\documentclass[12pt,a4paper]{article}
\pdfoutput=1
\usepackage{jheppub}  
\usepackage{amssymb} 
\usepackage{amsmath}
\usepackage{mathtools}
\usepackage{amsfonts}    
\usepackage{dsfont}
\usepackage{pdfpages}
\usepackage{verbatim}
\usepackage{tensor}
\usepackage{mathrsfs}
\usepackage[mathscr]{euscript}
\usepackage{tikz}
\usetikzlibrary{patterns,decorations.pathreplacing}

\newcommand{\ba}{\begin{align}}

\newcommand{\be}{\begin{equation}}
\newcommand{\ee}{\end{equation}}
\def\bd{\begin{tikzpicture}}
\def\ed{\end{tikzpicture}}

\newcommand{\abs}[1]{\left| #1 \right|}

\renewcommand\Im{\mathop{\text{Im}}}

\renewcommand\Re{\mathop{\text{Re}}}

\newcommand{\sphere}{\text{sphere}}
\newcommand{\torus}{\text{torus}}
\allowdisplaybreaks[1]

\title{Partition functions of the tensionless string}

\author{Lorenz Eberhardt} 
\affiliation{School of Natural Sciences, Institute for Advanced Study, \\
\hspace*{0.3cm}Einstein Drive 1, Princeton,  NJ 08540, USA}
\emailAdd{elorenz@ias.edu}

\abstract{We consider string theory on $\text{AdS}_3 \times \text{S}^3 \times \mathbb{T}^4$ in the tensionless limit, with one unit of NS-NS flux. This theory is conjectured to describe the symmetric product orbifold CFT. We consider the string on different Euclidean backgrounds such as thermal $\text{AdS}_3$, the BTZ black hole, conical defects and wormhole geometries. In simple examples we compute the full string partition function. We find it to be independent of the precise bulk geometry, but only dependent on the geometry of the conformal boundary. For example, the string partition function on thermal $\text{AdS}_3$ and the conical defect with a torus boundary is shown to agree, thus giving evidence for the equivalence of the tensionless string on these different background geometries. We also find that thermal $\text{AdS}_3$ and the BTZ black hole are dual descriptions and the vacuum of the BTZ black hole is mapped to a single long string winding many times asymptotically around thermal $\text{AdS}_3$. Thus the system yields a concrete example of the string-black hole transition.
Consequently, reproducing the boundary partition function does not require a sum over bulk geometries, but rather agrees with the string partition function on any bulk geometry with the appropriate boundary. We argue that the same mechanism can lead to a resolution of the factorization problem when geometries with disconnected boundaries are considered, since the connected and disconnected geometries give the same contribution and we do not have to include them separately.}

\begin{document}

\maketitle

\makeatletter
\g@addto@macro\bfseries{\boldmath}
\makeatother

\section{Introduction}

The $\text{AdS}/\text{CFT}$ correspondence \cite{Maldacena:1997re} is a strong-weak duality. Weakly coupled gauge theory probes the deeply quantum regime of the bulk, where the string becomes tensionless and the radius of $\text{AdS}$ is small in units of string length. Usually a description of such a bulk system is not directly accessible and bulk computations are often done in the supergravity limit. The tensionless limit is the regime where the AdS/CFT correspondence is `simplest' and is potentially provable, see \cite{Berkovits:2007rj, Berkovits:2019ulm} for direct attempts on $\text{AdS}_5 \times \text{S}^5$. Exposing the inner workings of the correspondence does not only teach us important lessons about the formulation of string perturbation theory, but also expose the physics of string theory in this regime that is somewhat removed from our geometric intuition. Finally, the tensionless limit roughly corresponds to the unhiggsed phase of string theory and thus exploring it teaches us something about the underlying symmetries of the string \cite{Gross:1987ar}.

$\text{AdS}_3$ is an arena where many technicalities simplify, but interesting physics still remains. In particular, the program of the tensionless limit can be completely carried out and equivalence of the string theory with the conjectured boundary CFT is essentially proven \cite{Gaberdiel:2018rqv, Giribet:2018ada, Eberhardt:2018ouy, Eberhardt:2019ywk, Eberhardt:2020akk, Hikida:2020kil}. For a brief overview see \cite{Giribet:2020mkc}. The precise statement is that IIB string theory on the background $\text{AdS}_3 \times \text{S}^3 \times \mathbb{T}^4$ supported by one unit of NS-NS flux is equivalent to the symmetric orbifold CFT $\text{Sym}^N(\mathbb{T}^4)$. The worldsheet theory can be described in terms of a WZW model based on the supergroup $\text{PSU}(1,1|2)$, which for one unit of flux becomes essentially a free theory.
This definition of tensionless string is somewhat different to previous attempts in the literature in that the level of the Kac-Moody symmetry on the worldsheet is not critical \cite{Karlhede:1986wb, Isberg:1993av, Gustafsson:1994kr, Lindstrom:2003mg, Bakas:2004jq}. The model is tensionless in the sense that is possesses massless higher spin fields in its spectrum.
\medskip

In the previous works mentioned above, the tensionless string was always considered on global $\text{AdS}_3$ $(\times \text{S}^3 \times \mathbb{T}^4)$. Many interesting aspects of the $\text{AdS}_3/\text{CFT}_2$ correspondence emerge only when we consider spaces with different conformal boundaries, such as thermal $\text{AdS}_3$ with a torus boundary. For instance, there are Euclidean black hole geometries: the BTZ black hole and all its $\mathrm{SL}(2,\mathbb{Z})$ images. In the classical gravitational way to evaluate the boundary torus partition function from the bulk, one should sum over all possible saddles of the gravitational theory and correct the result of these saddles by loop-corrections. This computation has a long history in the case of $\text{AdS}_3$, see e.g.~\cite{Maloney:2007ud, Yin:2007gv, Giombi:2008vd, Keller:2014xba, Benjamin:2019stq, Alday:2019vdr, Benjamin:2020mfz,Maxfield:2020ale}.\footnote{For very small radii, the dual CFT can become a RCFT and the partition function obtained this way matches on the nose \cite{Castro:2011zq, Jian:2019ubz}.}
For low temperature, thermal AdS is the dominating saddle, whereas for high temperature black holes dominate. Thus, this picture leads to Hawking-Page phase transitions in the boundary theory \cite{Hawking:1982dh, Keller:2011xi}.

There is a related gravitational computation that leads to a serious puzzle in the $\text{AdS}_3$/$\text{CFT}_2$ correspondence, to which we will refer to as the factorization problem. There are also saddles of the gravity theory that have disconnected boundaries, such as a wormhole connecting two genus 2 surfaces. From a CFT point of view, the partition function on the disjoint union of the two genus 2 surfaces is simply the product of the two individual partition functions. When computing the same quantity in gravity, we expect that disconnected as well as connected bulk geometries contribute. 
The basic problem is that while the disconnected contributions lead to a factorized answer, the factorization is destroyed by the connected contributions. The optimist's answer to the problem is that the connected geometries all cancel and factorization is restored. While this is an open question in 3d gravity, this does not happen in JT gravity, where the sum over geometries becomes a genus expansion \cite{Saad:2019lba}. Instead the correct boundary description is interpreted as an ensemble of theories. In this case, the partition function on two disconnected Riemann surfaces does not factorize because of the ensemble average \cite{Saad:2019lba, Stanford:2019vob}. The same is expected to happen in 3d gravity \cite{Cotler:2020ugk, Belin:2020hea} and was recently argued for in a toy example \cite{Afkhami-Jeddi:2020ezh, Maloney:2020nni}. The same conclusion is expected the hold generically from an EFT point of view for chaotic quantum systems \cite{Altland:2020ccq}.
This resolution is not an option for a stringy AdS/CFT correspondence because there is a unique conjectured dual and no natural ensemble over which to average is known. Furthermore, the density of states is discrete, as is appropriate for a single dual theory.
\medskip

In this paper we address these phenomena for the tensionless string on $\text{AdS}_3$. Both the Hawking-Page phase transition and the factorization problem have to be partially revised in this setting. We restrict to the somewhat simpler Euclidean setting. 

A recurring theme in the literature is the claim that a black hole of string size transitions into a single long string that winds around the horizon \cite{Susskind:1993ws, Horowitz:1996nw, Giveon:2005mi}. We find a picture that confirms this view very explicitly. The bulk theory does not contain black holes or rather the black hole background is equivalent to the long winding string in thermal $\text{AdS}_3$. 

We start by computing the string partition function on thermal $\text{AdS}_3$. Several special properties of the string on this background allow us to perform this computation \emph{exactly} in both $\alpha'$ and $g_\text{string}$ (under some favorable assumptions whose validity is discussed in Section~\ref{subsec:discussion}). A major input for this computation is the conjecture that the worldsheet moduli localize on holomorphic covering spaces of the boundary torus \cite{Pakman:2009zz, Eberhardt:2019ywk, Eberhardt:2020akk}. This in particular implies that the worldsheet computation is one-loop exact and that the integral over the moduli space of tori for the worldsheet one-loop contribution reduces to a discrete sum. We do not know much about non-perturbative corrections but they seem also absent. It is then surprising that the string spectrum on thermal $\text{AdS}_3$ accounts for the entire boundary partition function (up to choice of spin structure). It is not necessary to sum over different gravitational saddles to recover the boundary partition function. This seems in tension with the existence of the Hawking-Page transition, but we explain that it can be understood as a transition from the thermal $\text{AdS}_3$ vacuum to thermal $\text{AdS}_3$ with a long string winding around the boundary. This suggests that the black hole can be identified with the single long string. Even though the string itself is light for small string coupling, the winding number is of order $g_\text{string}^{-2}$, which makes it backreact on the geometry.
Thus there are (at least) two descriptions of the black hole: in terms of the classical Euclidean BTZ background or in terms of a highly excited string on thermal $\text{AdS}_3$. Hence there is a duality relating these two pictures. 
 This is a very explicit realization of the black hole-string transition. The computation of the partition function gives a very direct construction of the `black hole' microstates in this instance.

We solidify this picture by looking at another consistent string background with a torus boundary: the conical defect. In this case, we can also compute the string partition function exactly using the same technology, since the boundary is still a torus. We again find agreement of the string partition function with the full boundary torus partition function. However, the agreement is not trivial. Since again a large number of fundamental strings are involved, the sphere and the torus contributions to the string partition function can be of comparable size and the spectrum is completely rearranged compared to the thermal $\text{AdS}_3$. In particular, the vacuum of the conical defect with deficit angle $2\pi(1-M^{-1})$ gets mapped to many strings that wind $M$ times around the boundary of thermal $\text{AdS}_3$. This leads to a similar duality as for the black hole case.

While these are the only backgrounds on which we could reliably compute the string partition function, they teach us some important lessons that seem to carry over to more complicated backgrounds. Of particular interest to us is the wormhole that connects two genus 2 surfaces. We discuss that the computations with one torus boundary suggest that the connected and disconnected geometries are dual descriptions of each other and lead to the same factorized partition function, thus resolving the factorization problem we mentioned earlier. 
One intuitive way to understand these results is as follows. The single NS5-brane 
does not possess a throat and the strings do not fall inside the throat. Instead, they stay close to the boundary of $\text{AdS}_3$ and are insensitive to the internal features of the bulk manifolds. While this reasoning makes the proposed dualities sound trivial, changing the background reorders the states in intricate ways.

\medskip

This paper is organized as follows. We start in Section~\ref{sec:locally Euclidean AdS3} by reviewing consistent string backgrounds that are locally $\text{AdS}_3$. They can all be obtained by taking various orbifolds of global $\text{AdS}_3$. Section~\ref{sec:thermal AdS3 partition function} contains our main calculation of the string partition function on thermal $\text{AdS}_3$. We review the necessary background on the $\text{PSU}(1,1|2)_1$ WZW model, discuss the orbifold that reduces the theory to thermal $\text{AdS}_3$ and evaluate the one-loop partition function completely. It is convenient to perform this calculation in a grand canonical ensemble, which we shall discuss in detail. The sphere contribution to the partition function is somewhat subtle. At the end of the Section, we include a discussion of the various features of the calculation, such as the interpretation of the Hawking-Page transition and the black hole/string transition. We repeat a similar calculation in Section~\ref{sec:conical defect}, where we consider conical defect geometries. This calculation confirms and strengthens the conclusions gained in the case of thermal $\text{AdS}_3$. Finally, we extrapolate our findings in Section~\ref{sec:higher genus} to more complicated backgrounds. We discuss the generalization to higher genus boundaries and  the factorization problem. We end with a summary of the most important points, a discussion and future directions in Section~\ref{sec:discussion}.
\section{Locally Euclidean \texorpdfstring{$\text{AdS}_3$}{AdS3} spaces} \label{sec:locally Euclidean AdS3}
Let us start by recalling some basic facts about spaces that are locally Euclidean $\text{AdS}_3$. We often also denote Euclidean $\text{AdS}_3$ by $\mathbb{H}^3$ -- hyperbolic space.
 In string theory, we can consider also spaces with orbifold singularities, but let us first focus on the smooth geometries. These are by definition hyperbolic 3-manifolds. Hyperbolic 3-manifolds are incredible rich and there is a zoo of possibilities. The universal covering space of every hyperbolic 3-manifold is $\mathbb{H}^3$. Thus, every hyperbolic 3-manifold can be written as $\mathbb{H}^3/\Gamma$ for a discrete subgroup $\Gamma \subset \mathrm{PSL}(2,\mathbb{C})$ that acts properly discontinuously on $\mathbb{H}^3$ (this requirement is weakened if we allow for orbifold singularities). Such groups are called Kleinian groups, for an overview see e.g.~\cite{Maskit:2012}.

The conformal boundary of $\mathbb{H}^3$ is the Riemann sphere $\mathbb{CP}^1$ and $\mathrm{PSL}(2,\mathbb{C})$ acts by M\"obius transformations on it. 
While a Kleinian group $\Gamma$ acts properly discontinuously on $\mathbb{H}^3$, it typically does not on $\mathbb{CP}^1$. Let $\Omega$ be the maximal open set in $\mathbb{CP}^1$ on which it does act properly discontinuously. We shall in the following assume that $\Omega \ne \varnothing$. The complement $\Lambda(\Gamma)=\mathbb{CP}^1 \setminus \Omega$ is called the domain of discontinuity or the limit set. 

We can hence construct hyperbolic three manifolds $\mathbb{H}^3/\Gamma$ with conformal boundary $\Omega/\Gamma$. $\Omega/\Gamma$ is in general a collection of (possibly singular) Riemann surfaces. This hence gives a simple recipe for constructing the string worldsheet theory for strings on these hyperbolic 3-manifolds. $\mathbb{H}^3$ is described by the $\mathbb{H}^3$ conformal field theory and consequently $\mathbb{H}^3/\Gamma$ should be described by an orbifold CFT of the $\mathbb{H}^3$-model. We will mostly consider the tensionless limit of superstring theory on (Euclidean) $\text{AdS}_3 \times \text{S}^3 \times \mathbb{T}^4$. This theory is most conveniently described in terms of the so-called hybrid formalism \cite{Berkovits:1999im, Eberhardt:2018ouy}, where the target space of the six large dimensions is the supergroup $\text{PSU}(1,1|2)$.\footnote{This is the Lorentzian target space. The Euclidean target space is the supersymmetrization of $\mathbb{H}^3 \times \mathrm{S}^3$ for which we do not know an appropriate name. $\text{PSU}(1,1|2)$ is meant to denote the Euclidean target space in the following. \label{footnote:Euclidean AdS3}}

The non-linear sigma model on $\mathbb{H}^3$ has the following action:\footnote{For the Euclidean model, $\gamma$ and $\bar{\gamma}$ are complex conjugates of each other, whereas in the $\text{SL}(2,\mathbb{R})$ model, they are independent and real.}
\begin{align}
S_{\text{AdS}_3}&=\frac{k}{4\pi} \int \mathrm{d}^2 z \ \sqrt{g} \left( \partial \Phi \bar{\partial} \Phi+\bar{\partial }\gamma \partial \bar{\gamma} \mathrm{e}^{2\Phi}\right) \\
&\sim \frac{k}{4\pi} \int \mathrm{d}^2 z \ \sqrt{g} \left(\partial \Phi \bar{\partial} \Phi+\beta \bar{\partial}\gamma+\bar{\beta} \partial \bar{\gamma}-\beta \bar{\beta}\mathrm{e}^{-2\Phi}\right) \ , \label{eq:AdS3 action}
\end{align}
where in the second line we passed to a first-order form.\footnote{There is a linear dilaton that is generated at the quantum level \cite{Giveon:1998ns, deBoer:1998gyt}. In our discussion, it will not play a major role. } Here, $(\mathrm{e}^\Phi, \gamma, \bar{\gamma})$ correspond to the Poincar\'e coordinates on $\text{AdS}_3$.
The action of $\mathrm{PSL}(2,\mathbb{C})$ on these fields is relatively complicated, but simplifies dramatically in the regime $\Phi \to \infty$, which corresponds to the conformal boundary of $\mathbb{H}^3$. The model that describes the quotient space $\mathbb{H}^3/\Gamma$ is obtained by performing the appropriate field identifications.

In the following, we will review some important cases of this construction that play a major role in this article. We mostly discuss the bosonic case and mention the additional features of the supersymmetric construction.
\subsection{Single torus boundary} \label{subsec:single torus boundary}
A very important and well-known case is where the boundary of $\mathbb{H}^3/\Gamma$ is a single torus. In this case, a complete classification of the possible groups $\Gamma$ can be given:
\begin{enumerate}
\item \label{item:thermal AdS3} $\Gamma \cong \mathbb{Z}$. The generator $g$  of the infinite cyclic group can be diagonalized into the form
\be 
g=\mathop{\text{diag}}(\mathrm{e}^{\pi i t}, \mathrm{e}^{-\pi i t} )\ , \label{eq:thermal AdS3 orbifold action}
\ee
where $t$ is the modular parameter of the resulting torus.\footnote{We use $t$ instead of $\tau$, since $\tau$ will denote the worldsheet modular parameter.} The resulting 3-manifold is smooth and is depicted in Figure~\ref{fig:thermal AdS3}. Depending on which cycle in the torus is interpreted as time, the resulting geometry can have different physical interpretations. The 1-cycle of the torus is contractible in the bulk. Thus, interpreting the $1$-cycle as space and the $t$-cycle as time, the space cycle becomes contractible in the bulk, whereas the time cycle does not. This is thermal $\text{AdS}_3$. Reversing the roles of $1$ and $t$ leads to the Euclidean BTZ black hole. In general, we can declare any cycle $a+bt$ for $(a,b)=1$ to be the time direction in the torus, which leads to the well-known $\mathrm{SL}(2,\mathbb{Z})$ family of Euclidean black holes in $\text{AdS}_3$. 

\begin{figure}
\begin{center}
\includegraphics[width=.5\textwidth]{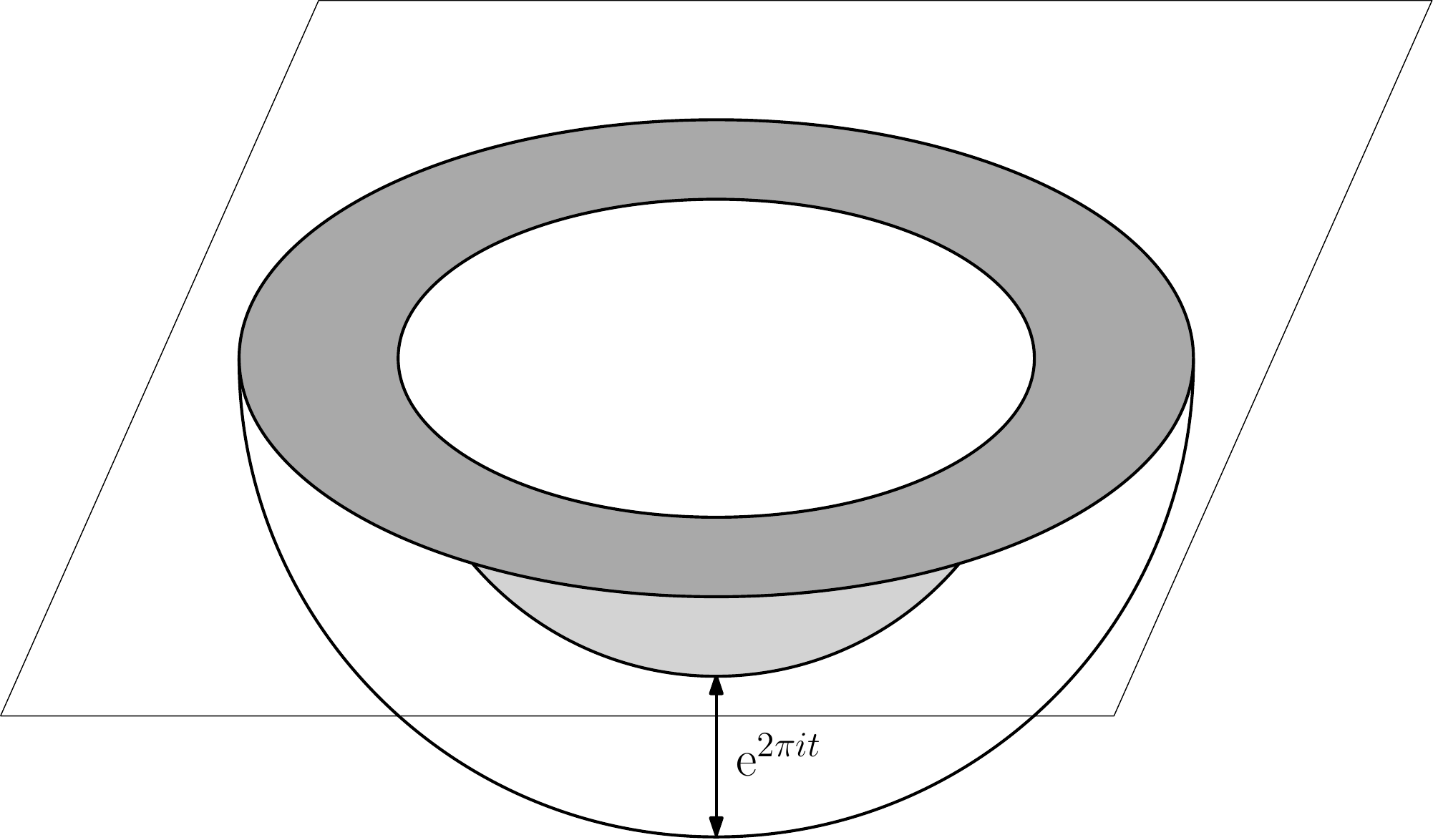}
\end{center}
\caption{Thermal $\text{AdS}_3$ obtained from the Schottky parametrization. Here we drew $\text{AdS}_3$ in Poincar\'e coordinates. The boundary is the plane indicated in the figure, while the bulk extends underneath the plane. The fundamental domain of the group action \eqref{eq:thermal AdS3 orbifold action} on the boundary is an annulus, which is depicted in dark gray. In the bulk, the orbifold action identifies the two hemispheres that extend from the inner and outer circle of the annulus. The resulting space is fibred by hemispheres that bound circles in the annulus. Thus, it is topologically a disk times a circle, i.e.\ a solid torus. The non-contractible cycle corresponds to the radial direction on the boundary, i.e.~to time in radial quantization.
} \label{fig:thermal AdS3}
\end{figure}

In the supersymmetric setting, we also have to specify a flat $\mathrm{SU}(2)$ bundle on the boundary torus, which we can implement by a Wilson line. Around the space cycle, the $\mathrm{SU}(2)$ bundle is periodic, but along the time cycle, we perform a rotation in the $\mathrm{SU}(2)$ bundle by a matrix
\be 
\mathop{\text{diag}}(\mathrm{e}^{\pi i z}, \mathrm{e}^{-\pi i z})\ . \label{eq:SU2 group element}
\ee
In total, the orbifold group is hence generated by the $\mathrm{PSU}(1,1|2)$ matrix
\be 
\mathop{\text{diag}}(\mathrm{e}^{\pi i t}, \mathrm{e}^{-\pi i t} ,\mathrm{e}^{\pi i z}, \mathrm{e}^{-\pi i z})\ ,\label{eq:thermal AdS3 supersymmetric orbifold action}
\ee
which generates a cyclic subgroup $\mathbb{Z}$ of $\mathrm{PSU}(1,1|2)$. The relevant supersymmetric thermal $\text{AdS}_3$ is then the quotient space $\mathrm{PSU}(1,1|2)/\mathbb{Z}$.
\item $\Gamma \cong \mathbb{Z} \times \mathbb{Z}_M$ for some integer $M \ge 2$ (in the case $M=1$, we again obtain the thermal $\text{AdS}_3$ geometry we discussed above). In this case, there are two generators $g$ and $h$, where $h$ is of finite order. We can diagonalize them simultaneously and they take the form
\be 
g=\mathop{\text{diag}}(\mathrm{e}^{\frac{\pi i t}{M}},
 \mathrm{e}^{-\frac{\pi i t}{M}} 
)\ , \qquad h=\mathop{\text{diag}}(
\mathrm{e}^{\frac{\pi i}{M}} , \mathrm{e}^{-\frac{\pi i}{M}}
)\ . \label{eq:conical defect orbifold action}
\ee
Thus, we are applying a further identification on the geometries discussed in the previous case. The resulting geometry is not smooth, but has a conical singularity in the center of the space. See Figure~\ref{fig:conical defect} for a schematic depiction. As in the previous case, we can still choose different time cycles on the boundary, which will lead to different physical interpretations. Even in the simplest case where we choose the $1$-cycle to be the space direction and the $t$-cycle to be the time direction, the space cycle is no longer contractible in the bulk, since it gets stuck on the conical singularity. We will refer to these spaces as the conical defect geometries.

\begin{figure}
\begin{center}
\includegraphics[width=.5\textwidth]{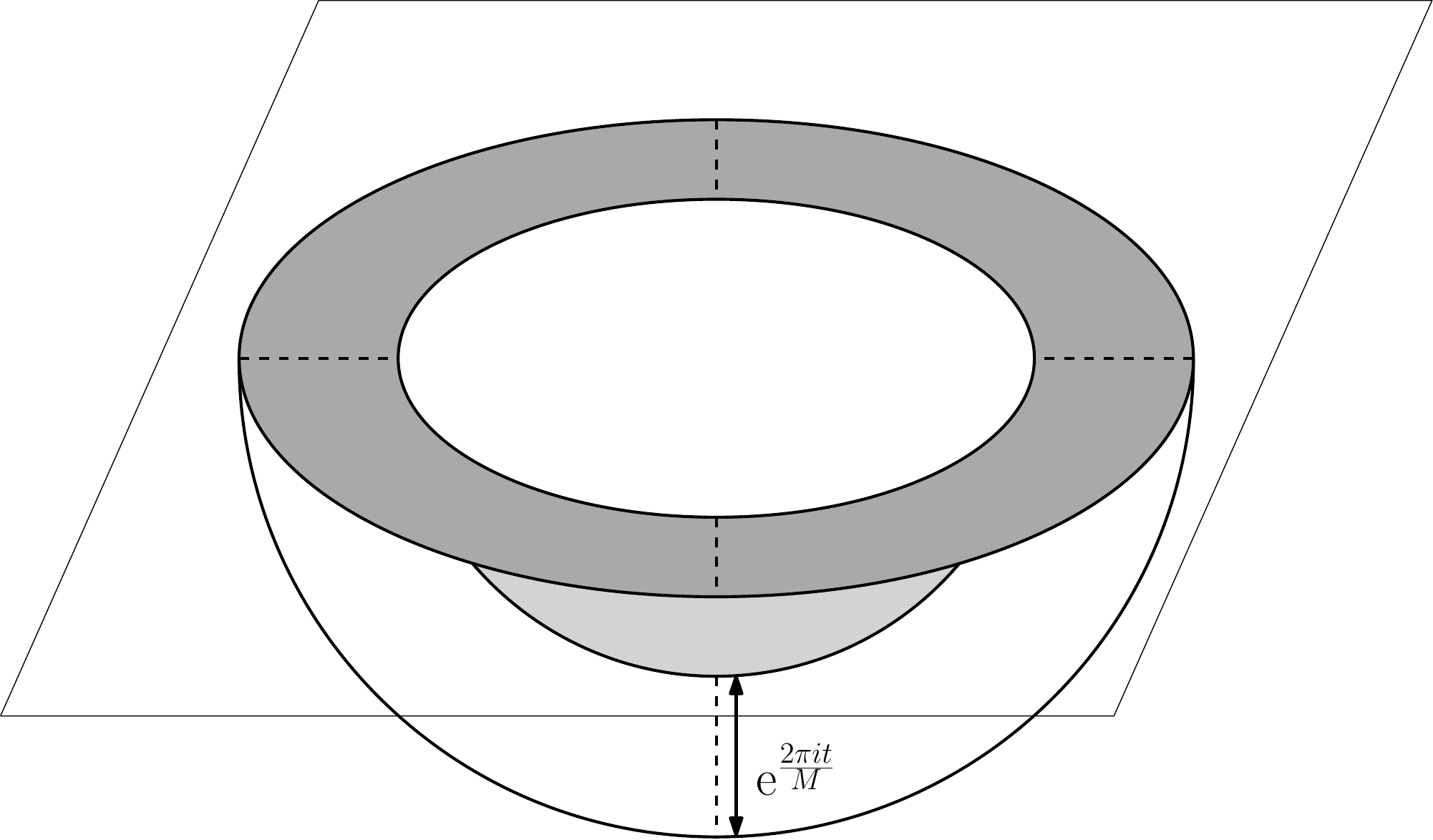}
\end{center}
\caption{Conical defect obtained from the Schottky parametrization. We chose $M=4$ in the figure. This is the same as for thermal $\text{AdS}_3$, except for the additional identification that reduces the fundamental domain to a slice of the annulus.} \label{fig:conical defect}
\end{figure}

There is again a supersymmetric analogue of these geometries. We do not want to introduce an $\mathrm{SU}(2)$ rotation for the $\mathrm{PSL}(2,\mathbb{C})$ element $h$, since this is an identification around the space cycle of the boundary torus. However, $g$ should again be combined with the $\mathrm{SU}(2)$ group element \eqref{eq:SU2 group element}. Thus the supersymmetric quotient group is generated by
\be 
\tilde{g}=\mathop{\text{diag}}(
\mathrm{e}^{\frac{\pi i t}{M}} ,
\mathrm{e}^{-\frac{\pi i t}{M}} , \mathrm{e}^{\pi i z} , \mathrm{e}^{-\pi i z}) \ , \quad
\tilde{h}=\mathop{\text{diag}}(
\mathrm{e}^{\frac{\pi i}{M}}, \mathrm{e}^{-\frac{\pi i}{M}} , 1 , 1
)\ \in \text{PSU}(1,1|2)\ .
\ee
\item $\Gamma \cong \mathbb{Z} \times \mathbb{Z}$. In this case there are two commuting generators which are both non-diagonalizable and can be taken to be
\be 
g_1=\begin{pmatrix}
1 & 1 \\ 0 & 1
\end{pmatrix}\ , \qquad
g_t=\begin{pmatrix}
1 & t \\ 0 & 1
\end{pmatrix}\ ,
\ee
where again $t$ is the modular parameter of the resulting torus. This geometry also has a singularity in the bulk, but not of conical type. One can think of this geometry as a cone over a torus. Since the generators of this orbifold are not diagonalizable but parabolic, it is much harder to describe this space in string theory. Thus, we will not consider it in this paper and focus on the first two.
\end{enumerate}
\subsection{Multiple torus boundaries} \label{subsec:multiple torus boundaries}
There are no hyperbolic 3-manifolds with multiple torus boundaries (not even singular hyperbolic manifolds). This is somewhat unfortunate for our purposes in this paper, since we would like to compute the string partition function on these backgrounds. Spaces with disconnected boundaries can be interpreted as Euclidean wormhole geometries and as discussed in the introduction are of particular physical interest.

This statement is easy to prove. The boundary $\bigsqcup_{i=1}^n \mathbb{T}^2_i$ can by definition be written as $\Omega/\Gamma$, where $\Omega$ is an open subset of $\mathbb{CP}^1$ and $\Gamma$ the corresponding Kleinian group. Since $\Omega$ is a covering space of the collection of tori, it follows that $\Gamma$ is a subgroup of $\mathbb{Z}^{2n}$ and is hence abelian. The set of abelian Kleinian groups is however very small and we essentially already saw all of them. It now follows quickly that there can at most be one torus boundary.

We should also note that the same argument also shows that there cannot be any hyperbolic 3-manifold with boundary spheres. The only space with a boundary sphere is $\mathbb{H}^3$. This also follows from the general theorem \cite{Witten:1999xp}.
\subsection{Single higher genus boundary} \label{subsec:single higher genus boundary}
Next, we consider a single genus $g \ge 2$ surface as a boundary. This is much more complicated than the $g=1$ case. There is a simple class of such hyperbolic three-manifolds, given by handlebodies. Handlebodies are obtained when $\Gamma$ is a Schottky group.\footnote{A Schottky group $\Gamma$ is a rank $g$ freely generated Kleinian group whose elements are all loxodromic. A $\text{PSL}(2,\mathbb{C})$ transformation $\gamma$ is called loxodromic if $\text{tr}\,(\gamma)^2 \not\in \mathbb{R}$. } This generalizes the construction of thermal $\text{AdS}_3$ to higher genus. There is a uniformization theorem that ensures that there is a Schottky group for any choice of moduli of the boundary surface. In the case of 3d gravity, this was analyzed in detail in \cite{Krasnov:2000zq, Yin:2007gv}.
However the story is much more complicated in this case and there are many non-handlebody hyperbolic 3-manifolds with a genus $g \ge 2$ boundary. We refer to \cite{Takhtajan:2002cc, Yin:2007at} for further details.
\subsection{Two higher genus boundaries}\label{subsec:two higher genus boundaries}
Finally, we consider geometries with two higher genus boundaries. These geometries can be obtained by choosing $\Gamma$ to be a Fuchsian group (or a quasi-Fuchsian group when the moduli of the two boundary surfaces are chosen independently). Fuchsian groups are discrete subgroups of $\mathrm{PSL}(2,\mathbb{R}) \subset \mathrm{PSL}(2,\mathbb{C})$. They act separately on the upper and lower half plane. Thus $\Omega$ has two components in this instance, which leads to the two boundary surfaces. Bers' simultaneous uniformization theorem ensures that for any choice of moduli of the two boundary surfaces, there is a quasi-Fuchsian group that realizes the wormhole geometry in between them \cite{Bers:1960}.

\section{The tensionless string on thermal \texorpdfstring{$\text{AdS}_3$}{AdS3}} \label{sec:thermal AdS3 partition function}
In this section we start with the simplest geometry discussed above -- supersymmetric thermal $\text{AdS}_3$.
We compute the tensionless string partition function on $\text{thermal AdS}_3 \times \mathrm{S}^3 \times \mathbb{T}^4$ where the theory is conjectured to be equivalent to the symmetric product orbifold \cite{Maldacena:1997re, Eberhardt:2018ouy, Eberhardt:2019ywk}. The theory on global $\text{AdS}_3$ is conveniently described in the hybrid formalism \cite{Berkovits:1999im}, which is built on the $\mathrm{PSU}(1,1|2)_k$ WZW model \cite{Gotz:2006qp}.\footnote{Since we are considering \emph{Euclidean} $\text{AdS}_3$, the real form of the WZW model is actually different. Our treatment is completely algebraic and the difference does not play a major role in the following. See also footnote~\ref{footnote:Euclidean AdS3}.} The tensionless limit is described by the value $k=1$. The $k=1$ model behaves very differently from higher $k$'s; in particular it does not contain short string states and instead of a continuum of long strings, only one particular energy is allowed and hence the resulting spectra are discrete.

In this section, we will first review the important features that are needed to compute the worldsheet partition function on global $\text{AdS}_3$ and will then discuss the orbifold that reduces the theory to thermal $\text{AdS}_3$. We give a physical interpretation of the results in Section~\ref{subsec:thermal AdS3 lessons}.
\subsection{The \texorpdfstring{$\mathrm{PSU}(1,1|2)_1$}{PSU(1,1|2)1} WZW model}
The relevant model for the tensionless limit is the $\mathrm{PSU}(1,1|2)_1$ WZW model, which was analyzed in detail in \cite{Eberhardt:2018ouy}. 
The reader who does not want to know all the technical details can jump to eq.~\eqref{eq:psu112 partition function}, which is the torus partition function of $\mathrm{PSU}(1,1|2)_1$.

The global subalgebra $\mathfrak{psu}(1,1|2)$ that is formed by the zero modes of the affine currents has only very few representations that are allowed in the affine algebra as highest weight representations. The essential reason for this is that the subalgebra $\mathfrak{su}(2)_1 \subset \mathfrak{psu}(1,1|2)_1$ has only the spin $0$ and spin $\frac{1}{2}$ representations as affine highest weights. In fact, the only allowed ground state representation of $\mathfrak{psu}(1,1|2)_1$ is denoted by $\mathscr{F}_\lambda$. Here, $\lambda\in \mathbb{R}/\mathbb{Z}$ describes the quantization of $J_0^3$, the Cartan generator of $\mathfrak{sl}(2,\mathbb{R})$. When decomposing this $\mathfrak{psu}(1,1|2)$ representation into the bosonic subalgebra $\mathfrak{sl}(2,\mathbb{R}) \oplus \mathfrak{su}(2)$, it takes the following form
\be 
\begin{tabular}{ccc}
& $(\mathscr{C}^{j=\frac{1}{2}}_\lambda,\mathbf{2})$ & \\
$(\mathscr{C}^{j=1}_{\lambda+\frac{1}{2}},\mathbf{1})$ & & $(\mathscr{C}^{j=0}_{\lambda+\frac{1}{2}},\mathbf{1})$
\end{tabular} 
\ee
Here, $\mathbf{m}$ denotes the $m$-dimensional $\mathfrak{su}(2)$ representation and $\mathscr{C}^{j}_\lambda$ denotes an $\mathfrak{sl}(2,\mathbb{R})$ representation in the continuous series with spin $j$ and $J_0^3$-quantization specified by $\lambda$. This representation is BPS and has vanishing quadratic Casimir.\footnote{When one thinks of $\mathrm{PSU}(1,1|2)$ as the global part of $\mathcal{N}=4$ superconformal symmetry, this is the $h=\frac{1}{2}$ BPS representation, provided one chooses the discrete representation $\mathscr{D}^j_j$ instead of the continuous representation $\mathscr{C}^j_\lambda$. Thus, the usual discrete BPS representation is a subrepresentation of this representation for $\lambda=\frac{1}{2}$.}

 In the full theory, this representation (or rather the affine representation build on top of this representation), as well as all its spectrally flowed images $\sigma^w(\mathscr{F}_\lambda)$ appear.\footnote{Also indecomposable modules appear, but on the level of the character, they are indistinguishable from these modules. This is related to the fact that for $\lambda=\frac{1}{2}$, the representation $\mathscr{F}_\lambda$ is no longer irreducible. For details, see \cite{Eberhardt:2018ouy}.} We follow \cite{Eberhardt:2018ouy} and abuse the notation by denoting from now on by $\mathscr{F}_\lambda$ the corresponding affine representation.
 Its character reads
 \begin{align} 
 \mathrm{ch}[\mathscr{F}_\lambda](\theta,\zeta;\tau)&=\text{tr}\left((-1)^{\text{F}} \mathrm{e}^{2\pi i \theta J_0^3}\mathrm{e}^{2\pi i \zeta K_0^3} \mathrm{e}^{2\pi i \tau(L_0+\frac{1}{12})}\right)\\
 &=\sum_{r \in \mathbb{Z}+\lambda} \mathrm{e}^{2\pi i r\theta}  \frac{\vartheta_1\big(\frac{\theta+\zeta}{2};\tau\big)\vartheta_1\big(\frac{\theta-\zeta}{2};\tau\big)}{\eta(\tau)^4}\ .
 \end{align}
 Here, we have as before introduced the chemical potential $\theta$ for $\mathfrak{sl}(2,\mathbb{R})$. We have similarly also introduced a chemical potential $\zeta$ for $\mathfrak{su}(2)$ (whose Cartan generator we denoted by $K_0^3$). The character includes a $(-1)^{\text{F}}$ to obtain good modular properties. The shift $+\tfrac{1}{12}$ is the usual shift $-\tfrac{c}{24}$, because $c=-2$ in this case. 
 The infinite sum leads to distributional characters, but as we shall see this is a crucial property of the model.
 The Hilbert space has the structure
 \be 
\mathcal{H} \cong  \bigoplus_{w \in \mathbb{Z}}\ \int\limits_{\mathbb{R}/\mathbb{Z}}\hspace{-.55cm}\boldsymbol{\oplus}\, \mathrm{d}\lambda \ \sigma^w \big(\mathscr{F}_\lambda\big) \otimes \overline{\sigma^w\big(\mathscr{F}_\lambda\big)}\ .
 \ee
 Note that this is quite different from the $\mathrm{SL}(2,\mathbb{R})_k$ WZW model (or the $\text{PSU}(1,1|2)_k$ WZW model for $k>1$), which on top of this involves an integral over all possible $\mathfrak{sl}(2,\mathbb{R})$ spins, as well as the inclusion of discrete representations.
 
 Thus, the $\mathrm{PSU}(1,1|2)_1$ partition function becomes
 \begin{multline} 
 Z_{\mathfrak{psu}(1,1|2)_1}(\theta,z,\tau)=\sum_{w \in \mathbb{Z}} \int_0^1 \mathrm{d}\lambda\ \sum_{r,\, \bar{r} \in \mathbb{Z}+\lambda} \mathrm{e}^{2\pi i r (\theta-w\tau)-2\pi i \bar{r}(\bar{\theta}-w \bar{\tau})} |q|^{w^2}\\
 \times\abs{ \frac{\vartheta_1\big(\frac{\theta+\zeta}{2};\tau\big)\vartheta_1\big(\frac{\theta-\zeta}{2};\tau\big)}{\eta(\tau)^4}}^2\ .
 \end{multline}
Let us look at the sum over $r$ and $\bar{r}$ in the sector $w=0$, together with the integral over $\lambda$. Denoting $s=r-\bar{r}$ and $\Delta=r+\bar{r}$, we can write this term as
\begin{align}
\frac{1}{2}\int_{-\infty}^\infty \mathrm{d}\Delta \sum_{s=-\infty}^\infty \mathrm{e}^{2\pi i s \Re \theta-2\pi \Delta \Im \theta}
\end{align}
Through analytic continuation, we can define this to be
\be 
\frac{1}{2}\delta_\mathbb{Z}^2(\theta) \equiv \frac{1}{2}\delta_{\mathbb{Z}}(\Re \theta) \delta(\Im \theta)=\frac{1}{2}\sum_{r \in \mathbb{Z}} \delta^2(\theta-r)\ . \label{eq:delta theta}
\ee
Thus, the partition function can be written as
\be 
Z_{\mathfrak{psu}(1,1|2)_1}(\theta,z;\tau)=\frac{1}{2}\sum_{r,w \in \mathbb{Z}} \delta^2(\theta-w \tau-r)|q|^{w^2} \abs{ \frac{\vartheta_1\big(\frac{\theta+\zeta}{2};\tau\big)\vartheta_1\big(\frac{\theta-\zeta}{2};\tau\big)}{\eta(\tau)^4}}^2\ . \label{eq:psu112 partition function}
\ee
This is the torus partition function of the Euclidean $\mathfrak{psu}(1,1|2)_1$ theory. The appearance of the $\delta$-functions seems somewhat formal, but we will later interpret them as a localization property of the model. 
The partition function is formally modular covariant and transforms as
 \be 
Z_{\mathfrak{psu}(1,1|2)_1} \left(\frac{\theta}{c\tau+d},\frac{\zeta}{c\tau+d}; \frac{a \tau+b}{c \tau+d} \right)= \abs{\exp\left(\frac{\pi i (\zeta^2-\theta^2) c}{2(c \tau+d)} \right)}^2 Z_{\mathfrak{psu}(1,1|2)_1}(\theta,\zeta;\tau)
\ee
for $\begin{pmatrix}
a & b \\ c & d
\end{pmatrix} \in \text{SL}(2,\mathbb{Z})$.
\subsection{The worldsheet partition function}
As a next step, we review the worldsheet partition function including both the contributions from the internal manifold and the ghosts. As discussed in \cite{Eberhardt:2018ouy}, the ghosts cancel effectively the oscillator contributions from $\mathfrak{psu}(1,1|2)_1$ and we end up with
\be 
Z_{\text{worldsheet, global AdS}_3}(\theta,\zeta;\tau)=\frac{1}{2}\sum_{r,w \in \mathbb{Z}} \delta^2(\theta-w \tau-r)|q|^{w^2} Z_{\tilde{\text{R}}}^{\mathbb{T}^4}(\theta,\zeta;\tau)\ . \label{eq:AdS3 partition function}
\ee
Here, $Z_{\tilde{\text{R}}}^{\mathbb{T}^4}(\theta,\zeta;\tau)$ is the R-sector partition function of the sigma model on $\mathbb{T}^4$ with insertions of $(-1)^\text{F}$. We have refined it by two chemical potentials that correspond to the $\mathfrak{so}(4)_1$ symmetry that the fermions generate. One of those will become the R-symmetry in the dual CFT. Explicitly,
\be 
Z_{\tilde{\text{R}}}^{\mathbb{T}^4}(\theta,\zeta;\tau)=\left|\frac{\vartheta_1(\frac{\theta-\zeta}{2};\tau)\vartheta_1(\frac{\theta+\zeta}{2};\tau)}{\eta(\tau)^6} \right|^2 \Theta^{\mathbb{T}^4}(\tau)\ ,
\ee
where $\Theta^{\mathbb{T}^4}(\tau)$ is the (non-holomorphic) theta function of the Narain lattice of the torus.
We can bring this formula in a more standard form by using that $\theta=w\tau+r$ inside the partition function of $\mathbb{T}^4$, thanks to the presence of the delta function. By spectral flow, we have
\be 
Z_{\tilde{\text{R}}}^{\mathbb{T}^4}(w\tau+r,\zeta;\tau)=|q|^{-\frac{w^2}{2}}Z^{\mathbb{T}^4} \begin{bmatrix}
\frac{r}{2} \\ \frac{w}{2}
\end{bmatrix}
(\zeta;\tau)\ .
\ee
Here, we wrote
\be 
Z^{\mathbb{T}^4}\begin{bmatrix}
0 \\
0
\end{bmatrix}=Z_{\tilde{\text{R}}}^{\mathbb{T}^4}\ , \quad
Z^{\mathbb{T}^4}\begin{bmatrix}
\frac{1}{2} \\
0
\end{bmatrix}=Z_{\text{R}}^{\mathbb{T}^4}\ , \quad
Z^{\mathbb{T}^4}\begin{bmatrix}
0 \\
\frac{1}{2} 
\end{bmatrix}=Z_{\tilde{\text{NS}}}^{\mathbb{T}^4}\ , \quad
Z^{\mathbb{T}^4}\begin{bmatrix}
\frac{1}{2} \\
\frac{1}{2}
\end{bmatrix}=Z_{\text{NS}}^{\mathbb{T}^4}\ .
\ee
The arguments in the parenthesis label the spin structures and take values in $\frac{1}{2}\mathbb{Z}/  \mathbb{Z}$. The two entries stand for the periodicity in the time and the space direction of the worldsheet torus. Under modular transformations, the spin structures with non-zero entries transform into each other, whereas the $\tilde{\text{R}}$ spin structure is invariant. We have
\be 
Z^{\mathbb{T}^4}\begin{bmatrix}
\mu \\
\nu
\end{bmatrix}\left(\frac{a \tau+b}{c \tau+d}\right)=Z^{\mathbb{T}^4}\begin{bmatrix}
d\mu+b \nu \\
c \mu+a \nu
\end{bmatrix}(\tau)\ .
\ee
We can thus write
\be 
Z_{\text{worldsheet, global AdS}_3}(\theta,\zeta;\tau)=\sum_{r,w \in \mathbb{Z}} Z\begin{pmatrix}
w & r
\end{pmatrix}(\theta,\zeta;\tau)
\ee
with
\be 
Z\begin{pmatrix}
w & r
\end{pmatrix}(\theta,\zeta;\tau)=\frac{1}{2} \delta^2(\theta-w \tau-r)\mathrm{e}^{-\pi w \Im \theta} Z\begin{bmatrix}
\frac{r}{2} \\ \frac{w}{2}
\end{bmatrix}(\zeta;\tau)\ . 
\ee
We again used the localization constraint to simplify the term $\abs{q}^{\frac{w^2}{2}}$.

Under modular transformations, $(w,r)$ behaves as a doublet. More precisely the transformation behaviour is
 \begin{multline} 
Z\begin{pmatrix}
w & r
\end{pmatrix} \left(\frac{\theta}{c\tau+d},\frac{\zeta}{c\tau+d}; \frac{a \tau+b}{c \tau+d} \right)\\
= |c \tau+d|^2\abs{\exp\left(\frac{\pi i (\zeta^2-\theta^2) c}{2(c \tau+d)} \right)}^2 Z\begin{pmatrix} aw+rc & bw+rd\end{pmatrix}(\theta,\zeta;\tau)\ . 
\end{multline}
The modular weight $(1,1)$ is expected for the worldsheet partition function and will be canceled against the weight of the measure $\frac{\mathrm{d}^2 \tau}{\Im \tau}$ when integrating over the fundamental domain.
\subsection{The orbifold partition function}
Next, we consider the orbifold partition function of this worldsheet theory. In the bosonic theory, this process is described in Appendix~\ref{app:bosonic}. Similar orbifolds corresponding to the supersymmetric conical orbifold were explored in \cite{Martinec:2001cf, Martinec:2002xq}.

The orbifold will have twisted sectors labelled by two integers $m$ and $n$, corresponding to the twists
\be 
\text{\scalebox{.7}{$m$}}\underset{n}{\raisebox{-5pt}{\text{\scalebox{2}{$\square$}}}}
\ee
along the cycles of the worldsheet torus.
 Thus, sectors of the theory are in total labelled by four integers -- $m$, $n$ and $w$, $r$ that we encountered already above. We will put them in a $2 \times 2$ matrix
\be 
M=\begin{pmatrix}
w & r \\ -n & m
\end{pmatrix}
\ee
Under modular transformations, $M$ transforms as $M\gamma^{-1}$ for $\gamma \in \mathrm{SL}(2,\mathbb{Z})$. For $(w,r)$ we have established this above and for $(m,n)$ this is just the usual orbifold transformation, see eq.~\eqref{eq:orbifold transformation}. 

We have
\be 
\text{\scalebox{.7}{$m$}}\underset{0}{\raisebox{-5pt}{\text{\scalebox{2}{$\square$}}}}= \Im t \ Z_{\text{AdS}_3}(mt,mz;\tau)\ ,
\ee
and hence
\be 
Z{\begin{pmatrix}
w & r \\ 0 & m
\end{pmatrix}}=\Im t \ Z\begin{pmatrix}
w & r
\end{pmatrix}(m t,mz;\tau)\ . \label{eq:thermal AdS3 untwisted sector}
\ee
This follows directly from the interpretation of the torus partition function as a trace over the Hilbert space and the action of the orbifold generator \eqref{eq:thermal AdS3 supersymmetric orbifold action}.
The prefactor $\Im t$ can be thought of as the inverse volume of the orbifold group. Its appearance is explained in more detail in Appendix~\ref{subapp:bosonic orbifold}.
Here, we generalized the notation of the previous section to account for all sectors described by the four integers. 

We should note that $t$ and $z$ are the parameters of the orbifold and are no longer interpreted as chemical potentials on the worldsheet, but as chemical potentials in spacetime. In particular, $t$ is interpreted as the modular parameter in spacetime. This means in particular that they are fixed under worldsheet modular transformations, contrary to the transformation behaviour of $\theta$ and $\zeta$, see eq.~\eqref{eq:psu112 partition function}.

Note that putting $z=0$ implies that the fermions are periodic around the spacetime thermal cycle and thus the partition function is computed with a fixed spacetime spin structure ($\tilde{\text{NS}}$). We will later consider also the antiperiodic spin structure (NS) that is then simply obtained by sending $z \to z+1$, since the fermions have half-integer R-charge.

Thus, we can deduce
\be 
Z\begin{pmatrix}
a w &b w \\ c m & dm
\end{pmatrix}=\frac{1}{|c \tau+d|^2} Z\begin{pmatrix}
w & 0 \\ 0 & m
\end{pmatrix}\left(\frac{a \tau+b}{c \tau+d} \right)
\ee
for $\begin{pmatrix}
a & b \\ c & d 
\end{pmatrix}\in \text{SL}(2,\mathbb{Z})$.
Since every integer $2 \times 2$ matrix can be written in this way, this gives the unique answer
\begin{multline} 
Z\begin{pmatrix}
a & b \\ c & d
\end{pmatrix}=\frac{1}{2} \Im t\,\delta^2\big(t(c \tau+d)-a \tau-b\big)\left| \exp\left(\frac{\pi i t}{2} \det\begin{pmatrix}
a & b \\ c & d
\end{pmatrix}+\frac{\pi i c(c \tau+d)}{2}z^2\right)\right|^2\\
\times Z^{\mathbb{T}^4}\begin{bmatrix}
\frac{b}{2} \\ \frac{a}{2}
\end{bmatrix}(z(c \tau+d);\tau)\ . \label{eq:Zabcd}
\end{multline}
The full worldsheet partition function is given by summing over all matrices,
\be 
Z_{\text{worldsheet, thermal AdS}_3}(t,z;\tau)=\sum_{a,\,b,\,c,\,d \,\in \mathbb{Z}}Z\begin{pmatrix}
a & b \\ c & d
\end{pmatrix}\ . \label{eq:worldsheet partition function}
\ee
Let us make some comments about this result:
\begin{enumerate}
\item The partition function localizes on configurations for which
\be 
(c\tau+d)t=a\tau+b\ .
\ee
As we will discuss below, if the modular parameters are related in this way, there is a holomorphic map from the worldsheet torus to the boundary torus of degree equal to the determinant of the matrix.
\item Modular invariance only relates integers $a,\,b,\,c,\,d$ with the same determinant. Thus, the factor
\be 
\exp\left(\frac{\pi it}{2} \, \det\begin{pmatrix}
a & b \\ c & d
\end{pmatrix}\right)
\ee
plays no role for modular invariance. We will compute the result below in a grand canonical ensemble, where the determinant will be interpreted as the number of strings in the background.  \label{item:modular orbits}
\item Charge conjugation reverses the sign of $(a,b,c,d)$. Since the theory is invariant under charge conjugation, we could half the summation range which would cancel the factor $\tfrac{1}{2}$. We have indeed
\be 
Z\begin{pmatrix}
a & b \\ c & d
\end{pmatrix}=
Z\begin{pmatrix}
-a & -b \\ -c & -d
\end{pmatrix}\ .
\ee
\end{enumerate}
\subsection{Geometric interpretation} \label{subsec:geometric interpretation}
Let us take stock and give a geometric interpretation for the integers $(a,b,c,d)$. They label the different sectors of the model and correspond to the homotopy classes of maps of the worldsheet into the asymptotic region of spacetime,
\be 
X: \mathbb{T}^2_\tau \longrightarrow \partial(\text{thermal AdS}_3) \cong \mathbb{T}^2_t\ .
\ee
Since $\pi_2(\mathbb{T}^2)=0$, it follows from general results that the set of homotopy classes is given by
\be 
\{(a,b) \in \pi_1(\mathbb{T}^2)^2\, |\, ab=ba\}\cong \mathbb{Z}^4\ ,
\ee
corresponding to the four integers $(a,b,c,d)$. In fact, these maps are even realized by holomorphic maps.\footnote{This follows immediately from the Riemann existence theorem.} Let 
\be 
\Gamma:\mathbb{T}^2_{\tau} \to \mathbb{T}^2_{t}\ . 
\ee
be a holomorphic map, where we indicated the respective modular parameters. Then we can lift it up to a map $\tilde{\Gamma}: \mathbb{C}^2 \longrightarrow \mathbb{C}^2$. Since the map $\tilde{\Gamma}$ has to map the lattice $\Lambda_\tau$ of the first torus to the second one, it has to be an affine linear map, $\tilde{\Gamma}(z)=\alpha z+\beta$ for $\alpha,\,\beta \in \mathbb{C}$. In order for this map to descend to the torus, we need $\tilde{\Gamma}(\Lambda_\tau) \subset \Lambda_t$. In particular, $\tilde{\Gamma}(0) \in \Lambda_t$ and $\tilde{\Gamma}(1) \in \Lambda_t$, which implies that $\alpha,\,\beta \in \Lambda_t$. Thus, it only remains to check that
\be 
\tilde{\Gamma}(\tau)=\alpha \tau+\beta \in \Lambda_t\ ,
\ee
and hence also $\alpha\tau \in \Lambda_t$. Thus, $\tau$ has to be the ratio of two lattice elements in $\Lambda_t$, $\tau=\frac{\gamma}{\alpha}$, $\alpha,\,\gamma\in \Lambda_t$. It follows conversely that 
\be 
t=\frac{a \tau+b}{c \tau+d}\ , \qquad \tau=\frac{d t-b}{-c t+a}
\ee
for four integers $a,b,c,d \in \mathbb{Z}$. This is exactly the localization condition that we saw in the partition function. Since both $\Im \tau>0$ and $\Im t>0$, this is impossible to satisfy unless
\be 
\det \begin{pmatrix}
a & b \\ c & d
\end{pmatrix}=ad-bc>0\ . \label{eq:det ge 0}
\ee
For every choice of $a,b,c,d$, there is a canonical map $\tilde{\Gamma}: \mathbb{T}_\tau \longrightarrow \mathbb{T}_t$ that fixes the origin of the torus given by\footnote{The translation has no geometrically interesting meaning and does not change the homotopy class.} \footnote{Every torus has a $\mathbb{Z}_2$ automorphism and hence also $-\Gamma(z)$ is a possible map.}
\be 
\Gamma(z)=(a-c t) z\ .
\ee
The degree of this map is
\be 
\deg \Gamma=\det \begin{pmatrix}
a & b \\ c & d
\end{pmatrix}=ad-bc\ .
\ee
Let us now give a rough geometric picture what the different integers mean. We think of $t$ as being fixed, since it specifies the background geometry. First, we perform a modular transformation
\be 
\tau=\frac{d \tau'-\frac{b}{\det}}{-c \tau'+\frac{a}{\det}}\ ,
\ee
where $\det=ad-bc$, which gives then $t=\det \tau'$ and the corresponding canonical map becomes $\Gamma(z)=\det z$. Thus, up to modular transformations, the worldsheet just winds $ad-bc$ times asymptotically around $\text{AdS}_3$. 

In \cite{Eberhardt:2019ywk}, it was argued that this map $\Gamma$ can be directly identified with the worldsheet $\gamma$ that appears in the action \eqref{eq:AdS3 action} (or rather the one-point function of $\gamma(z)$ on the worldsheet). While in general, $\gamma$ is not holomorphic, it becomes holomorphic for $k=1$.

Let us also note in passing that the one-loop string partition function on \emph{global} $\text{AdS}_3$ vanishes, since it corresponds to \eqref{eq:AdS3 partition function} with $\theta=\zeta=0$. This is consistent with the fact that there is no holomorphic covering map $\mathbb{T}^2 \to \mathbb{CP}^1$.\footnote{Presumably the worldsheet sphere partition function should be non-zero for global $\text{AdS}_3$, similar to what is discussed below for the thermal $\text{AdS}_3$ case.} 
\subsection{String partition function}\label{subsec:thermal AdS3 string partition function}
We now evaluate the one-loop string partition function.\footnote{We thank Shota Komatsu for useful conversations surrounding the sphere partition function and the grand canonical ensemble.}

\paragraph{Grand canonical ensemble.}
String theory can have a variable number of fundamental strings in the geometry and from the point of view of the boundary, we are hence computing the partition function in the grand canonical ensemble where different values of $N$ (the number of copies in the symmetric product orbifold) are allowed, but we fix the corresponding chemical potential. This chemical potential can be incorporated into the action \eqref{eq:AdS3 action} by adding the following `topological term'
\be 
S_{\text{AdS}_3}'=S_{\text{AdS}_3}-2\pi i \sigma \int \gamma^* \omega\ .
\ee
Here $\omega$ is the volume form of the boundary surface of the hyperbolic 3-manifold, normalized such that $\int_{\partial \mathcal{M}_3} \omega=1$.\footnote{If $\partial \mathcal{M}_3$ is disconnected, we require $\int \omega=1$ for each boundary component. The orientation of the boundary and the worldsheet is induced from the complex structure.}
This term makes only sense for $k=1$, where we can identify $\gamma$ with the covering map from the worldsheet to the boundary. We have
\be 
\sigma \int_\Sigma  \gamma^* \omega=\sigma \deg(\gamma) \int_{\gamma(\Sigma)} \omega=\sigma  \deg(\gamma)\ ,
\ee
which shows that this term is indeed topological in this setting. While this term has a clear classical interpretation, it is a bit ad hoc. The same term can be introduced directly on the worldsheet, where it corresponds to the spacetime identity operator \cite{Porrati:2015eha, Kutasov:1999xu}. It was a puzzle that the spacetime identity operator is proportional to the winding number $w_i$. This puzzle is resolved once one fixes the chemical potential associated to the identity (or computes in the canonical ensemble where the number of strings is fixed). 
The grand canonical ensemble corresponds physically to fixing the boundary value of the $B$-field, whereas in the canonical ensemble the electric flux is fixed.

Thus, the path integral will weight a string configuration by $p^{\deg(\gamma)}$, where $p=\mathrm{e}^{2\pi i\sigma}$. As we discussed above,
\be 
\deg(\gamma)=\deg(\Gamma)=\det \begin{pmatrix}
a & b \\ c & d
\end{pmatrix}\ ,
\ee
and we already noticed in point \ref{item:modular orbits} above that we can insert $p^{\deg(\gamma)}$ in the string one-loop partition function while keeping modular invariance intact.

From a dual CFT point of view, we are computing the following partition function
\be 
\mathfrak{Z}(z,t,\sigma)=\sum_{N=0}^\infty p^N \mathcal{Z}_{\text{Sym}^N(\mathbb{T}^4)}(z,t)\ ,
\ee
where $p=\mathrm{e}^{2\pi i \sigma}$ is the chemical potential, as on the worldsheet.

In the context of the elliptic genus of the symmetric orbifold CFT (considering as seed theory K3), it is standard to consider the grand canonical ensemble, which leads to Siegel modular forms.

\paragraph{String path integral.} Before going on, let us briefly recall the string path integral. The general form of the string partition function takes the form
\be 
\mathcal{Z}=\exp\left(\frac{1}{g_\text{string}^2} \sphere+\torus+ \dots \right) \label{eq:string partition function}
\ee
For the one-loop partition function, we do not include higher genera. The exponential includes also disconnected worldsheets. The expression in the exponential is the string free energy.
Usually, we evaluate the path integral at fixed $g_\text{string}$ and arbitrary number of strings in the background. This is however not quite what we want in the present context. We have $g_\text{string}^{-2} \sim N$ under the holographic dictionary, where $N$ represents the number of fundamental strings in the background. As we have discussed above, $N$ for a single string is to be identified with $\deg(\Gamma)=ad-bc$, which is unbounded.\footnote{While it would be consistent from the point of view of modular invariance to bound $N$ on the worldsheet, see point~\ref{item:modular orbits} above, it is not consistent with the worldsheet fusion rules that are discussed in \cite{Eberhardt:2018ouy, Eberhardt:2019ywk}.} Thus, when describing the system in string perturbation theory, we necessarily need to include arbitrarily large winding sectors. Put differently, within the framework of string perturbation theory, there cannot be a worldsheet theory that describes the symmetric orbifold for fixed $N$.

We can however still compute the string partition function at fixed chemical potential $\sigma$ in the grand canonical ensemble. In this context, this is the natural ensemble of string perturbation theory. The grand canonical one-loop string partition function takes the form
\be 
\mathfrak{Z}_\text{string}(z,t,\sigma)=\exp\left(\sum_{N=1}^\infty \left(p\, \mathrm{e}^{\sphere}\right)^N\torus^{(N)}\right)\ ,
\ee
where the superscript means that we restrict the partition function to the $N$-wound sector. This formula can be derived as follows. Starting with the standard string partition function \eqref{eq:string partition function}, we can factor out the sphere contribution and obtain
\be 
\mathcal{Z}=\mathrm{e}^{N \text{sphere}} \exp\left(\torus\right)\ .
\ee
The sphere does not contribute to the number of strings in the background, since the worldsheet remains small and does not reach the boundary. Write $\torus=\sum_{n=1}^\infty \torus^{(n)}$. Then we can expand the exponential as follows:
\be 
\exp\left(\sum_{n=1}^\infty \torus^{(n)}\right)=\sum_{m_1,m_2,\dots,} \prod_{n\ge 1}\frac{1}{m_n!} \left(\torus^{(n)}\right)^{m_n}\ .
\ee
To obtain the canonical partition function, we have to fix the total degree of the torus contribution, i.e.\ $\sum_n m_n n=N$,
\be 
\mathcal{Z}_{\text{canonical},N}=\mathrm{e}^{N \text{sphere}} \sum_{\sum_n m_n n =N}\ \prod_{n\ge 1}\frac{1}{m_n!} \left(\torus^{(n)}\right)^{m_n}\ .
\ee
It is now straightforward to pass to a grand canonical ensemble,
\begin{align}
\mathfrak{Z}&=\sum_{N=0}^\infty p^N \mathcal{Z}_{\text{canonical},N} \\
&=\sum_{N=0}^\infty\left(p\,\mathrm{e}^{\text{sphere}}\right)^N \sum_{\sum_n m_n n =N}\ \prod_{n\ge 1}\frac{1}{m_n!} \left(\torus^{(n)}\right)^{m_n}\\
&=\sum_{m_1,m_2,\dots}\ \prod_{n\ge 1}\frac{\left(p\,\mathrm{e}^{\text{sphere}}\right)^{m_nn}}{m_n!} \left(\torus^{(n)}\right)^{m_n}\\
&=\exp\left(\sum_{N=1}^\infty \left(p\, \mathrm{e}^{\sphere}\right)^N\torus^{(N)}\right)\ .
\end{align}

\paragraph{Sphere contribution.} To compute the string partition function, we should start with the leading contribution -- the sphere partition function.
The sphere is hard to evaluate in usual string perturbation theory, because we still have to divide by the global conformal group $\mathrm{SL}(2,\mathbb{C})$, which naively gives 0. However, as usual, the worldsheet answer is expected to diverge because the target space has infinite volume. When properly regularized, one can obtain a finite answer, as was demonstrated for the two point function in \cite{Erbin:2019uiz}, and argued for in the case of $\text{AdS}_3$ in \cite{Kutasov:1999xu}. As far as we are aware, no direct computation of the sphere partition function has been performed.

Usually, the sphere contribution to the partition function is interpreted as the on-shell \emph{spacetime} action \cite{Maldacena:1998bw}.  This becomes problematic in our setting, since the spacetime theory is highly non-local and the supergravity action should not be trusted. We do not know how to properly compute the sphere contribution to the string partition function.\footnote{The spacetime on-shell action is proportional to the regularized volume of thermal $\text{AdS}_3$. This gives the answer $|q|^{-\frac{k}{2}}$, which when naively continued to $k=1$ gives the correct answer for a single asymptotic string. }

The sphere contribution is however fixed entirely by conformal invariance in the dual CFT and corresponds to the usual $|q|^{-\frac{c}{12}}$ factor in the partition function. 
In the present case, the central charge equals $c=6\times \text{number of strings}$ \cite{Brown:1986nw, Giveon:1998ns}. It is important to realize that a single string can lead to different contributions of the boundary central charge, depending on the number of times it asymptotically winds around $\mathrm{AdS}_3$. This can be derived from the worldsheet by a DDF-like construction \cite{Giveon:1998ns, Eberhardt:2019qcl}.\footnote{This construction is done for global $\text{AdS}_3$. The corresponding construction for the BTZ black hole was discussed in \cite{Rangamani:2007fz}.} This winding number is geometrically given by the degree of the map $\Gamma$ that we discussed in the previous subsection, which equals $ad-bc$. Thus, the sphere contribution is given by $\mathrm{e}^{\deg(\Gamma)\pi \Im t}$. In other words, to account for the sphere contribution, we propose to perform the replacement
\be 
\sigma \longmapsto \sigma-\frac{i}{2} \Im t\label{eq:sphere sigma replacement}
\ee
in the torus partition function.
This factor exactly cancels the opposite factor present in \eqref{eq:Zabcd}. 

\paragraph{Torus partition function.}
Because of the appearance of the $\delta$-function, we can evaluate the torus partition function in closed form. We have to integrate the worldsheet partition function \eqref{eq:worldsheet partition function} over the standard fundamental domain of the torus.
Since $(c,d)$ transform as a doublet under the modular group, we can extend for $(c,d) \ne (0,0)$ the region of integration from the fundamental domain to the vertical strip $-\frac{1}{2}\le  \Re \tau <\frac{1}{2}$ and restrict the summation over $c$ and $d$ to $c=0$ and $d\in \mathbb{Z}$. For $(c,d)=(0,0)$, the $\delta$-function reduces to $\delta^2(a \tau+b)$, which is impossible to satisfy inside the fundamental domain of the torus, except for $a=b=0$. If $a=b=0$, the $\delta$-function is always true. However, for $(c,d)=(0,0)$, the torus partition function is $Z_{\tilde{\text{R}}}^{\mathbb{T}^4}(0,0;\tau)$, which vanishes due to the fermion zero modes. Thus, the term $(c,d)=0$ does in fact not contribute.\footnote{Formally the $a=b=c=d=0$ term is $\delta^2(0)  Z_{\tilde{\text{R}}}^{\mathbb{T}^4}(0;\tau)$, which is ill-defined. Let us regulate it by choosing $d$ to be slightly non-zero, which leads to $\delta^2(d t) Z_{\tilde{\text{R}}}^{\mathbb{T}^4}(dz;\tau)$. This vanishes in the limit $d \to 0$.
If we would do the same analysis for K3, we would instead get $Z_{\tilde{\text{R}}}^{\mathbb{T}^4}(0;\tau)=24$, the Euler characteristic of K3. In this case, this would lead to a divergent contribution.} This is the same trick as was employed in \cite{Polchinski:1985zf, Maldacena:2000kv}.
It is slightly surprising that the string partition function is completely free from divergences.

After setting $c=0$, the partition function is still symmetric under charge conjugation $(a,b,d) \to (-a,-b,-d)$ and thus we can further restrict $d>0$ (remember that we already excluded the case $(c,d)=(0,0)$). We obtain the following connected string partition function
\begin{align}
\mathfrak{Z}_\text{conn}(t,z)&=\sum_{N=0}^\infty \int_{\mathcal{F}} \frac{\mathrm{d}^2 \tau}{\Im \tau} \ Z_{\text{worldsheet, thermal AdS}_3}^{(N)}(t,z;\tau)\mathrm{e}^{\pi N\Im t} p^{N} \\
&= \int_\mathcal{F}\frac{\mathrm{d}^2 \tau}{\Im \tau} \ \sum_{a,b,c,d \in \mathbb{Z}} \mathrm{e}^{\pi (ad-bc)\Im t} p^{ad-bc} Z \begin{pmatrix}
a & b \\ c & d
\end{pmatrix}\\ 
&=\Im t\int\limits_{-\frac{1}{2}<\Re \tau \le \frac{1}{2}} \frac{\mathrm{d}^2 \tau}{\Im \tau} \sum_{a,b \in \mathbb{Z}}\sum_{d=1}^\infty \delta^2(dt-a \tau-b)p^{ad}Z^{\mathbb{T}^4}\begin{bmatrix}
\frac{b}{2} \\ \frac{a}{2}
\end{bmatrix}(dz;\tau)\ .
\end{align}
Here, $Z_{\text{worldsheet}}^{(N)}$ denotes the $N=ad-bc$ sector of the worldsheet theory. In this formula, we already included the shift \eqref{eq:sphere sigma replacement}.
Since both $\Im \tau>0$, $\Im t>0$, only terms $a \ge 1$ can contribute in the above sum. This is the remnant of the condition \eqref{eq:det ge 0}.
Next, we can explicitly perform the integral,
\begin{align}
\mathfrak{Z}_\text{conn}(t,z)&=\Im t\int\limits_{-\frac{1}{2}<\Re \tau \le \frac{1}{2}} \frac{\mathrm{d}^2 \tau}{\Im \tau} \sum_{a,d=1}^\infty \sum_{b \in \mathbb{Z}}\delta^2(dt-a \tau-b)p^{ad} Z^{\mathbb{T}^4}\begin{bmatrix}
\frac{b}{2} \\ \frac{a}{2}
\end{bmatrix}(dz;\tau)\\
&= \frac{\Im t}{\Im \tau} \sum_{a,d=1}^\infty \frac{1}{a^2}\sum_{b \in \mathbb{Z}/a \mathbb{Z}}p^{ad} Z^{\mathbb{T}^4}\begin{bmatrix}
\frac{b}{2} \\ \frac{a}{2}
\end{bmatrix}(dz;\tau)\Bigg|_{\tau=\frac{d t+b}{a}}\ .
\end{align}
Since $\tau$ covers the vertical strip in the integration, $a \tau$ covers a $a$-fold wider strip and hence there are $a$ possible values for $b$ for which the localization constraint is satisfied. The factor $a^{-2}$ comes from the Jacobian of the $\delta$-function. Simplifying further, we finally obtain
\be 
\mathfrak{Z}_\text{conn}(t,z)= \sum_{a, d=1}^\infty \sum_{b \in \mathbb{Z}/a \mathbb{Z}} \frac{p^{ad}}{a d}  Z^{\mathbb{T}^4}\begin{bmatrix}
\frac{b}{2} \\ \frac{a}{2}
\end{bmatrix}\left(dz;\frac{d t+b}{a}\right)
\ee

The full disconnected one-loop partition function is now easy to obtain, we just have to exponentiate the connected contribution,
\be 
\mathfrak{Z}_\text{thermal AdS$_3$}(t,z,\sigma)=\exp\left[ \sum_{a, d=1}^\infty \sum_{b \in \mathbb{Z}/a \mathbb{Z}} \frac{p^{ad}}{ad}  Z^{\mathbb{T}^4} \begin{bmatrix}
\frac{b}{2} \\
\frac{a}{2}
\end{bmatrix}\left(dz;\frac{d t+b}{a}\right)\right]\ .
\ee
Geometrically, it is sometimes more illuminating to rewrite this as
\be
\mathfrak{Z}_\text{thermal AdS$_3$}(t,z,\sigma)=\exp\left[ \sum_{L=1}^\infty  \frac{p^{L}}{L}  T_L Z^{\mathbb{T}^4} \left(z;t\right)\right]\ ,
\ee
where $T_L $ is the (supersymmetric)  Hecke operator
\be 
T_L Z^{\mathbb{T}^4}(z;t)=\sum_{a\, |\, L} \sum_{b \in \mathbb{Z}/a \mathbb{Z}} Z^{\mathbb{T}^4} \begin{bmatrix}
\frac{b}{2} \\
\frac{a}{2}
\end{bmatrix}\left(dz;\frac{L t+a b}{a^2}\right)\ .
\ee
Here, $L$ is the degree of the covering map and the Hecke operator sums the partition function over the connected covering spaces of degree $L$.
\subsection{Comparison to the symmetric product orbifold partition function}
Let us compare this result to the torus partition function of the symmetric product orbifold. It is simplest to compute the $\tilde{\text{R}}$ sector partition function of the symmetric product orbifold \cite{Dijkgraaf:1996xw,Maldacena:1999bp}, which takes the form
\be 
\mathfrak{Z}_{\tilde{\text{R}}}^\text{Sym}(t,z,\sigma)=\exp\left[ \sum_{m, w=1}^\infty \sum_{r \in \mathbb{Z}/w \mathbb{Z}} \frac{p^{m w}}{m w}  Z^{\mathbb{T}^4} \begin{bmatrix}
0 \\
0
\end{bmatrix}\left(mz;\frac{m t+r}{w}\right)\right]\ .
\ee
Here, $w$ labels the twisted sectors and $r$ achieves the orbifold projection, which simply projects to all the states with $h-\bar{h} \in \mathbb{Z}$. The other three spin structures can be found by flowing $z \to z+1$ (R), $z \to z+t$ ($\tilde{\text{NS}}$) and $z \to z+t+1$ (NS). This gives
\begin{subequations}
\begin{align}
\mathfrak{Z}_{\text{R}}^\text{Sym}(t,z,\sigma)&=\exp\left[ \sum_{m, w=1}^\infty \sum_{r \in \mathbb{Z}/w \mathbb{Z}} \frac{p^{m w}}{m w}  Z^{\mathbb{T}^4} \begin{bmatrix}
\frac{m}{2} \\
0
\end{bmatrix}\left(mz;\frac{m t+r}{w}\right)\right]\ , \\
\mathfrak{Z}_{\tilde{\text{NS}}}^\text{Sym}(t,z,\sigma)&=\exp\left[ \sum_{m, w=1}^\infty \sum_{r \in \mathbb{Z}/w \mathbb{Z}} \frac{p^{m w}}{m w}  Z^{\mathbb{T}^4} \begin{bmatrix}
\frac{r}{2} \\
\frac{w}{2}
\end{bmatrix}\left(mz;\frac{m t+r}{w}\right)\right]\ , \label{eq:Sym NSt partition function}\\
\mathfrak{Z}_{\text{NS}}^\text{Sym}(t,z,\sigma)&=\exp\left[ \sum_{m, w=1}^\infty \sum_{r \in \mathbb{Z}/w \mathbb{Z}} \frac{p^{m w}}{m w}  Z^{\mathbb{T}^4} \begin{bmatrix}
\frac{r+m}{2} \\
\frac{w}{2}
\end{bmatrix}\left(mz;\frac{m t+r}{w}\right)\right]\ .\label{eq:Sym NS partition function}
\end{align}
\end{subequations}
By comparison with the string result, we have hence exactly
\be 
\mathfrak{Z}_{\text{thermal AdS}_3}=\mathfrak{Z}_{\tilde{\text{NS}}}^{\text{Sym}}\ .
\ee
This is the expected spin structure from thermal $\text{AdS}_3$. In general, when fixing a bulk manifold (as we did for thermal $\text{AdS}_3$), we should sum over all possible compatible spin structures. Since in thermal $\text{AdS}_3$ the spatial cycle is contractible, only the NS and $\tilde{\text{NS}}$ spin structures extend in the bulk.

We implicitly computed the supersymmetric partition function with $(-1)^\text{F}$ insertions. We can also compute the partition function in the $\text{NS}$ sector without $(-1)^\text{F}$ insertions, which corresponds to the other compatible spin structure on thermal $\text{AdS}_3$. 
This can easily be achieved by flowing $z \to z+1$, which exchanges the two spin structures. This is because $z$ had the interpretation of an $\mathrm{SU}(2)$ rotation by $\mathop{\text{diag}}(\mathrm{e}^{\pi i z},\mathrm{e}^{-\pi i z})$ around the spacetime thermal circle. Since fermions have half-integer charges, sending $z \to z+1$ changes their periodicity.
This flow can directly be performed on the worldsheet and since $z$ was just a parameter of the orbifold does not break worldsheet modularity.
This leads to the immediate matching between the $\text{NS}$ and the $\tilde{\text{NS}}$ spin structure and thermal $\text{AdS}_3$ with the corresponding spin structure.

As was already mentioned, the R-sector spin structure is not compatible with thermal $\text{AdS}_3$. Thus, the R-sector cannot correspond to thermal $\text{AdS}_3$. Since the S-modular transformation exchanges the $\tilde{\text{NS}}$ sector and the R-sector, the spatial cycle in the corresponding bulk geometry becomes non-contractible. This is the well-known Euclidean BTZ black hole.
One can question whether the distinction between the BTZ black hole and thermal $\text{AdS}_3$ makes sense in the tensionless limit. We will stick to the conventional geometric interpretation and consider them as two distinct bulk geometries.

\subsection{Lessons} \label{subsec:thermal AdS3 lessons}
We have seen that the thermal $\text{AdS}_3$ one-loop string partition function is already modular invariant (up to the choice of spin structure that break the modular group down to an index 3 subgroup) and accounts for the full symmetric orbifold partition function of the given spin structure. This is surprising for a variety of reasons.
\paragraph{One-loop exactness.}
We have only computed the one-loop partition function on thermal $\text{AdS}_3$. One would expect that also higher genus corrections are necessary to obtain a fully modular invariant partition function and presumably even D-instanton corrections. We will argue in Section~\ref{sec:higher genus} that at least the perturbative corrections vanish.
\paragraph{Sum over bulk geometries.} From a gravity point of view, we should sum over all possible bulk geometries in order to reproduce the boundary partition function \cite{Maloney:2007ud, Yin:2007gv, Giombi:2008vd, Keller:2014xba, Benjamin:2019stq, Alday:2019vdr, Benjamin:2020mfz,Maxfield:2020ale}. In this case, this involves all the $\mathrm{SL}(2,\mathbb{Z})$ images of thermal AdS and possibly also conical defects. 
The fact that the partition function is already modular invariant means that all the $\mathrm{SL}(2,\mathbb{Z})$ images  of thermal $\text{AdS}_3$ with the same spin structure have the same partition function which suggests that they are indistinguishable by string theory. We will argue that this picture holds much more generally for the tensionless string. Similarly, we will check in the next section that the conical defect also leads to the same partition function.
\paragraph{Hawking-Page phase transition.} For fixed (large) $N$, the symmetric orbifold possesses a Hawking-Page phase transition at temperature $T_\text{HP}=\frac{1}{2\pi}$ corresponding to the self-dual torus \cite{Hawking:1982dh, Keller:2011xi}. Conventionally this is interpreted as follows in the dual gravitational theory. The gravitational path integral instructs us to sum over all possible geometries compatible with the boundary, i.e.~at least the full $\mathrm{SL}(2,\mathbb{Z})$ family of black holes.
Below the Hawking-Page temperature $T_\text{HP}=\frac{1}{2\pi}$ (and at vanishing spin potential), this sum is dominated by thermal $\text{AdS}_3$, whereas above the Hawking-Page temperature the BTZ black hole becomes the dominating saddle. From this point of view, it is very counterintuitive that the entire CFT partition function arises from one bulk geometry. Of course, we still have to sum over spin structures, but this does not resolve the problem, since we can for example consider the $\text{NS}$ spin structure that is compatible both with thermal $\text{AdS}_3$ and the BTZ black hole. In our setting, the contribution to the NS spin structure from thermal $\text{AdS}_3$ and the BTZ black hole is identical. 

A semiclassical explanation of this apparent paradox is the following. Instead of summing over bulk geometries, we sum over a variety of string configurations, as encoded by the appearance of the $\delta$-functions in the worldsheet partition function \eqref{eq:worldsheet partition function}. The gravitational sum over saddles is replaced by the sum over worldsheet configurations. Naively, one would think that this is always dominated by the sphere topology in the large $N$ limit.
However, as discussed in Section~\ref{subsec:thermal AdS3 string partition function}, the torus partition function depends also on $N$, which is why one naturally computes the grand canonical partition function. Depending on the temperature, the torus contribution can compete with the sphere. Roughly speaking, for low temperatures, the sphere (or the vacuum) dominates, whereas the torus dominates for high temperatures.

Let us consider the case where $t$ is purely imaginary, i.e.~no spin potential is turned on.
Then there are two possible dominating string configurations. This parallels essentially the discussion for the symmetric product orbifold \cite{Hartman:2014oaa, Haehl:2014yla}. For low temperatures, the vacuum sector is dominating.
The vacuum of the dual CFT is identified with a gas of spherical worldsheets, together with toroidal worldsheets whose modular parameter agrees with the boundary modular parameter (i.e.~the terms with $a=d=\pm 1$ and $b=c=0$ in \eqref{eq:Zabcd}).

For high temperature, there is still a gas of spherical worldsheets, but only one toroidal worldsheet with modular parameter
\be 
\tau=-\frac{N}{t}\ .
\ee
This corresponds to the term $a=d=0$, $b=\pm N$ and $c=\mp 1$ in \eqref{eq:Zabcd}. Geometrically, this is one string that winds $N$ times around the boundary of thermal $\text{AdS}_3$.

\paragraph{The black hole/string transition.}
This gives another interpretation why we do not have to sum over different geometries. The contribution of the BTZ black hole to the symmetric orbifold partition function is replaced by a single perturbative very long string winding around the boundary of thermal $\text{AdS}_3$. This is in stark contrast to the situation for higher amounts of NS-NS flux, where black holes are expected to contribute to the dual CFT partition function. Thus, the tensionless string gives a concrete realization of the black hole/string transition \cite{Holzhey:1991bx, Susskind:1993ws, Horowitz:1996nw, Giveon:2005mi, Sen:1995in}. Let us recall the basic reasoning of this transition. If strings fall toward the horizon of a black hole they stretch more and more along the horizon. From the outside observer, the strings never reach the horizon. The entropy of a single string becomes so large that the thermodynamically most advantageous configuration is a single string.  This motivates the conjecture that string-sized black holes describe the same degrees of freedom as perturbative strings. In particular it was shown in \cite{Horowitz:1996nw} that their entropy agrees to leading order (and even logarithmic corrections in the extremal case \cite{Sen:1995in}).

Here we have seen this phenomenon for the tensionless string. The appearance of this transition for $k=1$ was already qualitatively discussed in \cite{Giveon:2005mi}. There are other controllable settings that are expected to exhibit the black hole/string transition such as the two-dimensional Euclidean black hole $\text{SL}(2,\mathbb{R})/\mathrm{U}(1)$ \cite{Witten:1991yr, Kutasov:2000jp, Nakayama:2005pk}. 
We have considered the Euclidean setting which has no horizon. We expect that the Lorentzian geometry is much richer and provides an ideal playground to explore this transition.

\paragraph{Hagedorn temperature.} Usually, perturbative string theory breaks down at the Hagedorn temperature $T_\text{Hagedorn}$ \cite{Atick:1988si}. This manifests itself in various ways. The main physical reason is that there is a new tachyon appearing from the winding of the thermal circle and thus the string one-loop partition function diverges. For $\text{AdS}_3\times \text{S}^3 \times \mathbb{T}^4$, the Hagedorn temperature is \cite{Berkooz:2007fe, Lin:2007gi}
\be 
T_\text{Hagedorn}=\frac{1}{2\pi} \sqrt{\frac{k(k+2)}{4k-1}}\ ,
\ee
which for $k=1$ agrees with the Hawking-Page temperature $T_\text{HP}=\frac{1}{2\pi}$. One might think that our computation should only be well-defined for $\Im t >1$, which corresponds to temperatures below the Hagedorn temperature.

What saves our calculation is again the grand canonical potential. From the point of view of the dual CFT, the Hagedorn temperature is a large $N$ effect and the partition function only diverges at infinite $N$. There is a reflection of this in the grand canonical ensemble. Consider the NS-sector partition function of the symmetric product orbifold \eqref{eq:Sym NS partition function}. The other spin structures can be discussed similarly, but lead to less stringent constraints. We should make sure that the sum over $m$ and $w$ converges, since a divergence would precisely signal a breakdown of string perturbation theory. Let us consider the case where both $\sigma$ and $t$ are purely imaginary. We first consider low temperature, i.e.~large $\Im t$. Then the dominating term comes from the vacuum of the partition function $Z^{\mathbb{T}^4}$ and the term $w=1$. These terms lead to the sum
\be 
\Im t \to \infty\,:\quad \mathfrak{Z}_{\tilde{\text{NS}}}^{\text{Sym}}(t,z,\sigma)\sim \exp\left[\sum_{m=1}^\infty \frac{1}{m}\mathrm{e}^{2\pi i m \sigma+\pi m \Im t} \right]\ ,
\ee
which converges, provided that
\be 
\Im t<2 \Im \sigma\ .
\ee
There is also a constraint from high temperature, which follows from the S-modular transformation in $t$. Together, we have
\be 
\Im t\,,\ \frac{1}{\Im t}<2 \Im \sigma\ .
\ee
We believe that these conditions are necessary and sufficient to ensure convergence. For chiral CFTs, this can be established more rigorously, see \cite{deLange:2018mri}.

Thus, the Hagedorn transition in the grand canonical ensemble appears as a critical chemical potential, but as long as we assume that the imaginary part of the chemical potential is big enough, there is no divergence and string perturbation theory should be trustworthy.

\section{Conical defect geometries}\label{sec:conical defect}
Let us move on to the next torus boundary geometry discussed in Section~\ref{subsec:single torus boundary} -- the conical defect geometry. We may get it from global $\text{AdS}_3$ by a $\mathbb{Z} \times \mathbb{Z}_M$ orbifold. The goal of this section is to compute the tensionless string partition function on this geometry. The main take away from this section is that the full string partition function is the \emph{same} as for thermal $\text{AdS}_3$ (that van be viewed as the conical defect with $M=1$). We will be somewhat briefer than for the thermal $\text{AdS}_3$, since the computation is very similar to that case.

The orbifold action \eqref{eq:conical defect orbifold action} is diagonal and we can thus implement the untwisted sector by the replacement $\theta \mapsto \frac{dt+n}{M}$, where $n \in \mathbb{Z}_N$ and $d \in \mathbb{Z}$. On the $\mathfrak{su}(2)$ chemical potential, we perform the replacement $\zeta \to dz$, which performs the same R-symmetry rotation around the non-contractible cycle as in the thermal $\text{AdS}_3$ case.

There is one subtlety we have to take care of. Since the spacetime spin structure is the $\tilde{\text{NS}}$ spin structure, we do not actually want to enforce periodicity in $t \mapsto t+1$, but rather in $t \mapsto t+2$ and thus the untwisted sector is obtained from the replacement $\theta \mapsto \frac{dt+2n}{M}$.\footnote{Or alternatively periodicity in simultaneous $t \mapsto t+1$ and $z \to z+1$.} This orbifold will be only consistent with the spin structure provided that $M$ is odd. Since the result is antiperiodic in $t \to t+1$ it would otherwise imply periodic boundary conditions in $t \to t+M$, which would imply that we would need to start with a $\tilde{\text{R}}$ spin structure. Thus, we will assume $M$ to be odd in the following.
\subsection{Worldsheet torus partition function}
The different sectors of the partition function are labeled by a matrix
\be 
\begin{pmatrix}
a & b \\ m & n \\ c & d
\end{pmatrix}\ ,
\ee
where as before $a,b,c,d \in \mathbb{Z}$ and additionally $m,\,n \in \mathbb{Z}_M$. This matrix transforms as three doublets under the modular group. The untwisted sector of the torus partition function of the orbifold is straightforward to write down. It is a simple generalization of \eqref{eq:thermal AdS3 untwisted sector} and reads
\be 
Z \begin{pmatrix}
a & b \\ 0 & n \\ 0 & d
\end{pmatrix}(t,z;\tau)=\frac{\Im t}{2M^2} \delta^2\left(\frac{dt+2n}{M}-a\tau-b\right)  \mathrm{e}^{-\frac{\pi ad \Im t}{M}} Z^{\mathbb{T}^4}\begin{bmatrix}
\frac{b}{2} \\ \frac{a}{2}
\end{bmatrix}(dz;\tau)\ . \label{eq:conical defect untwisted sector}
\ee
The additional prefactor $M^{-2}$ compared to the previous case comes about as follows. One power $M^{-1}$ is the volume of the group $\mathbb{Z}_M$. The volume of the group $\mathbb{Z}$ is also bigger by a factor of $M$, since we effectively replaced $t$ by $\frac{t}{M}$ in the orbifold action \eqref{eq:conical defect orbifold action} compared to \eqref{eq:thermal AdS3 orbifold action}. 
We perform a modular transformation to deduce
\begin{align}
Z& \begin{pmatrix}
a\alpha +b \gamma & a\beta+b\delta \\ n\gamma & n\delta \\ d\gamma & d\delta
\end{pmatrix}(t,z;\tau)\nonumber\\
&=\frac{1}{|\gamma\tau+\delta|^2}Z \begin{pmatrix}
a& b \\ 0 & n \\ 0 & d
\end{pmatrix}\left(t,z;\frac{\alpha\tau+\beta}{\gamma\tau+\delta}\right) \\
&=\frac{\mathrm{e}^{-\frac{\pi  ad \Im t}{M}} \Im t}{2M^2|\gamma\tau+\delta|^2} \delta^2 \left(\frac{dt+2n}{M}-a\frac{\alpha\tau+\beta}{\gamma\tau+\delta}-b\right)  Z^{\mathbb{T}^4}\begin{bmatrix}
\frac{b}{2}\\ \frac{a}{2}\end{bmatrix} \left(dz; \frac{\alpha \tau+\beta}{\gamma\tau+\delta}\right)\\
&=\frac{\mathrm{e}^{-\frac{\pi  ad \Im t}{M}}\Im t}{2} \delta^2 \big((dt+2n)(\gamma\tau+\delta)-aM(\alpha\tau+\beta)-bM(\gamma\tau+\delta)\big)\nonumber\\
&\qquad\times \left| \exp\left(\frac{\gamma\pi i}{2}(\gamma \tau+\delta) d^2z^2\right)\right|^2 Z^{\mathbb{T}^4}\begin{bmatrix}
\frac{1}{2}(a \beta+b\delta) \\ \frac{1}{2}(a\alpha+b\gamma)\end{bmatrix}(dz(\gamma \tau+\delta);\tau)\ .
\end{align}
We can almost uniquely read off the value for the different sectors.\footnote{The orbifold group has the non-trivial cohomology group $\mathrm{H}^2(\mathbb{Z} \times \mathbb{Z}_M;\mathrm{U}(1)) \cong \mathbb{Z}_M$ and thus one could introduce discrete torsion \cite{Vafa:1986wx, Vafa:1994rv}. The unique cocycle that can be introduced in the partition function is $\varepsilon((c,m),(d,n))=\mathrm{e}^{\frac{2\pi i}{M}x(cn-md)}$ for $x \in \mathbb{Z}_M$. We will continue with the model without discrete torsion and comment further in footnote~\ref{footnote:discrete torsion partition function} on the model with discrete torsion.}
We find
\begin{multline}
Z\begin{pmatrix}
a& b \\ m & n \\ c & d
\end{pmatrix}=\frac{N\Im t}{2} \delta^2 \big(tN(c \tau+d)-(aN-2m)\tau-(bN-2n)\big)\\
 \times\mathrm{e}^{-\pi  ((aN-2m)d-(bN-2n)c) \frac{\Im t}{N^2}}\left| \exp\left(\frac{c\pi i}{2}(c \tau+d) z^2\right)\right|^2 Z^{\mathbb{T}^4}\begin{bmatrix}
\frac{b}{2}\\ \frac{a}{2}\end{bmatrix}(z(c \tau+d);\tau)\ . \label{eq:Zabmncd}
\end{multline} 
Notice also that
\be 
\frac{1}{2}(aM-2m)=\frac{1}{2}a \bmod \mathbb{Z}\ , \qquad \frac{1}{2}(bM-2n)=\frac{1}{2}b \bmod \mathbb{Z}\ ,
\ee
because of our assumption that $M$ is odd.
Thus, we can define the combination $A=aM-2m$ and $B=bM-2n$, which allows us to write the result in a simple form,
\begin{multline}
Z\begin{pmatrix}
A & B \\ c & d
\end{pmatrix}= \frac{\Im t}{2} \delta^2 \big(t(c \tau+d)-A\tau-B\big)\exp\left(-\pi \det\begin{pmatrix}
A & B\\ c & d 
\end{pmatrix} \frac{\Im t}{M^2}\right)\\
\times  \left| \exp\left(\frac{c\pi i}{2}(c \tau+d) z^2\right)\right|^2 Z^{\mathbb{T}^4}\begin{bmatrix}
\frac{B}{2}\\ \frac{A}{2}\end{bmatrix}(z(c \tau+d);\tau)\ . \label{eq:ZABcd conical defect}
\end{multline}
Since $M$ is odd, $A$ and $B$ run exactly once over the integers.
At this point, the result looks almost identical to the result obtained for thermal $\text{AdS}_3$, compare to eq.~\eqref{eq:Zabcd} The only difference is the factor of $M^{-2}$ in the exponential. The full worldsheet partition is obtained by summing over all integers $A,B,c,d$.

We should note that the worldsheet partition function again localizes on configurations for which
\be 
t=\frac{A\tau+B}{c\tau+d}\ ,
\ee
which as discussed in Section~\ref{subsec:geometric interpretation} are the modular parameters for which there is a holomorphic covering map $\Gamma$ from the worldsheet to the boundary. This interpretation is unchanged and in particular
\be 
\deg(\Gamma)=\det\begin{pmatrix}
A & B \\ c & d
\end{pmatrix}
\ee
yields again the number of fundamental strings in the background.

\subsection{String partition function}
After having computed the worldsheet torus partition function, we can again compute the one-loop spacetime partition function. This follows the same steps as in Section~\ref{subsec:thermal AdS3 string partition function}.
\paragraph{Sphere contribution.} We again do not know how to compute the sphere partition function directly in the worldsheet theory. Let us first compute the value that is suggested by supergravity (even though we should not trust supergravity in this regime). The supergravity action simply computes the regularized volume of the resulting orbifold geometry. We have, see e.g.~\cite{Benjamin:2020mfz},\footnote{This result is intuitively straightforward to understand. The volume of $\mathbb{H}^3$ can be obtained from the volume of $\mathrm{S}^3$ by analytic continuation, which leads to $\mathop{\text{vol}}(\mathbb{H}^3)=-2\pi^2$. We should divide this by the volume of the orbifold group, which, as explained below eq.~\eqref{eq:conical defect untwisted sector}, equals $\frac{2M^2}{\Im t}$.}
\be 
\text{vol}(\mathbb{H}^3/(\mathbb{Z} \times \mathbb{Z}_M))=-\frac{\pi^2 \Im t}{M^2}\ ,
\ee
which suggests that the sphere contribution equals $\frac{1}{M^2}$ times the sphere contribution of thermal $\text{AdS}_3$, i.e. 
\be
\sphere=\det\begin{pmatrix}
a & b \\ c &d
\end{pmatrix}\frac{\pi \Im t}{M^2}\ , \label{eq:conical defect sphere contribution}
\ee
thus exactly cancelling the ground state energy in \eqref{eq:ZABcd conical defect}.
This can again be implemented by a shift in the chemical potential,
\be 
\sigma \longmapsto \sigma-\frac{i}{2M^2} \Im t\ . \label{eq:sphere sigma replacement conical defect}
\ee
Alternatively, this ground state energy may also be fixed by conformal invariance in the boundary CFT alone. Under the additional $\mathbb{Z}_M$ orbifold, the Virasoro generators in the dual CFT transform as
\be 
L_n \to \mathrm{e}^{\frac{2\pi i n}{M}} L_n\ .
\ee
Thus, the conical orbifold preserves only Virasoro generators $L_n$ with $n \in M\mathbb{Z}$. In order to bring the algebra back into a canonical form, we define
\be 
\tilde{L}_n=\frac{1}{M}L_{nM}+\frac{\tilde{c}}{24}\left(1-\frac{1}{M^2}\right)\ , \label{eq:conical defect Virasoro generators}
\ee
where $\tilde{c}=cM$. Then $\tilde{L}_n$ satisfies the Virasoro algebra with central charge $\tilde{c}=cM$. In the conical defect background, $\tilde{L}_n$ are the Virasoro generators. Thus the total (holomorphic) ground state energy should be
\be 
-\frac{\tilde{c}}{24}+\frac{\tilde{c}}{24}\left(1-\frac{1}{M^2}\right)=-\frac{\tilde{c}}{24 M^2}\ .
\ee
The first term is the ground state energy in thermal $\text{AdS}_3$ and the second the shift appearing in \eqref{eq:conical defect Virasoro generators}.
The central charge is $\tilde{c}=6 \times \text{number of strings}=6(ad-bc)$, which shows again that the sphere contribution should precisely cancel the ground state energy in \eqref{eq:ZABcd conical defect}.
\paragraph{One-loop partition function.} With these preparations, the string one-loop partition function in the conical defect geometry reads
\begin{align}
\mathfrak{Z}_\text{conn}(t,z,\sigma)&=\sum_{N=0}^\infty \int_{\mathcal{F}} \frac{\mathrm{d}^2 \tau}{\Im \tau} Z_\text{worldsheet, conical defect}^{(N)}(t,z;\tau) p^N \exp\left(N\frac{\pi \Im t}{M^2}\right)\\
 &=\int_{\mathcal{F}} \sum_{a,b,c,d \in \mathbb{Z}} p^{ad-bc} \frac{\Im t}{2} \delta^2 \big(t(c \tau+d)-a\tau-b\big)\nonumber\\
&\qquad\qquad\times \left| \exp\left(\frac{c\pi i}{2}(c \tau+d) z^2\right)\right|^2 Z^{\mathbb{T}^4}\begin{bmatrix}
\frac{b}{2}\\ \frac{a}{2}\end{bmatrix}(z(c \tau+d);\tau)\ . 
\end{align}
In this formula, we again renamed $A \to a$ and $B \to b$.
From here on, the calculation is identical to the thermal AdS$_3$ computation and we obtain\footnote{With the discrete torsion turned on, there are extra phases in the partition function. They can be summarized as
\be 
\mathfrak{Z}_\text{conical defect}^x(t,z,\sigma)=\mathfrak{Z}_{\tilde{\text{NS}}}^\text{Sym}(t,z,\sigma+\tfrac{x}{M})\ ,
\ee
where $x \in \mathbb{Z}_M$. In the canonical partition function, we would hence simply multiply the contribution from $\text{Sym}^N(\mathbb{T}^4)$ by $\mathrm{e}^{\frac{2\pi i x N}{M}}$. Thus, the canonical partition function only changes by a phase. We do not know an interpretation of this phase from the boundary point of view. \label{footnote:discrete torsion partition function} }
\be 
\mathfrak{Z}_\text{conical defect}(t,z,\sigma)=\mathfrak{Z}_{\text{thermal AdS}_3}(t,z,\sigma)=\mathfrak{Z}_{\tilde{\text{NS}}}^\text{Sym}(t,z,\sigma)\ .
\ee
We have thus given strong evidence for the claim that string propagation on thermal $\text{AdS}_3$ and the conical defect geometries is \emph{equivalent} in the tensionless limit.

In the conical defect geometries both cycles of the torus are non-contractible and thus there is geometrically no problem in putting any spin structure on the boundary. However it is not possible to obtain the R-sector spin structures by taking an orbifold of global $\text{AdS}_3$.
\subsection{Rearrangement of the spectrum} \label{subsec:conical defect rearrangement}
The conclusion that the string partition function on the conical defect geometry equals the string partition function on thermal $\text{AdS}_3$ might seem counter intuitive to the reader familiar with the dictionary of $\text{AdS}_3/\text{CFT}_2$. In \cite{Lunin:2002bj}, the gravity dual of twisted sector ground states in the symmetric product orbifold was established. 

Let us first consider a single-particle chiral primary in the symmetric orbifold, which means that it is a state in the twisted sector specified by the cyclic permutation $(1 \cdots M)$. 
Every twisted sector possesses 16 chiral primaries, which we may denote by $\sigma^{\varepsilon_1 \varepsilon_2 \varepsilon_3 \varepsilon_4}_M$, where $\varepsilon_i= \pm 1$. The conformal weight of these chiral primaries is
\be 
h=\frac{M}{2}+\frac{\varepsilon_1+\varepsilon_2}{4}\ , \qquad \bar{h}=\frac{M}{2}+\frac{\varepsilon_3+\varepsilon_4}{4}\ .
\ee
Let us focus on the ground state $\sigma^{----}_M$. 
In supergravity, this state can be interpreted as a coming from an $\mathrm{S}^3$ harmonic of spin $\frac{M-1}{2}$. For fixed $M$, the mass of this state is of order one and is hence not heavy enough to backreact on the geometry. This changes if we consider the twisted sector of the conjugacy class
\be 
(1 \cdots M)(M+1 \cdots 2M) \cdots ((n-1)M+1 \cdots nM)\ , \label{eq:conical defect state in symmetric product}
\ee
where we assume for simplicity that $nM=N$. This yields a (multiparticle) chiral primary of conformal weight
\be 
h=\bar{h}=\frac{n(M-1)}{2}=\frac{N(M-1)}{2M}
\ee
in the spectrum. This is of order $N$ and one can hence interpret it as a different bulk geometry. Lunin, Mathur and Saxena showed that the corresponding bulk geometry is the conical defect with deficit angle $2\pi(1-M^{-1})$, which is exactly the geometry we have considered in this section. Since we have placed the conical defect only in $\text{AdS}_3$, we should actually talk about the twisted sector ground state and not the chiral primary in this twisted sector.

These statements fit naturally together when we notice that while the total string spectrum agrees on the conical defect and thermal $\text{AdS}_3$, the orbifold rearranged the spectrum quite non-trivially. The untwisted sector of the $\mathbb{Z}_M$ orbifold corresponds to sectors with $m=0$ and $c=0$ in \eqref{eq:Zabmncd}, which translates in \eqref{eq:ZABcd conical defect} to sectors where $M \, |\, A$ and $c=0$. We thus see that $M\, |\, Ad-Bc$ and so the degree of the covering map is always a multiple of $M$. The untwisted sector of the orbifold hence maps only to twisted sectors whose twist is a multiple of $M$. In particular, the vacuum of the conical defect (i.e.~the state in the untwisted sector of the orbifold with the lowest spacetime energy) gets identified with the state \eqref{eq:conical defect state in symmetric product}, which is exactly the statement of \cite{Lunin:2002bj}. The missing twisted sectors of the symmetric orbifold arise from the perspective of the conical defect from the twisted sectors of the worldsheet orbifold. 

We should also note another phenomenon. The ground state energy of the $M$-twisted sector in thermal $\text{AdS}_3$ arises in the two backgrounds partially from the sphere and the torus, but the proportions are different. We already explained that the two topologies can contribute at the same order, since the $N^{-1}$ suppression of the torus can be compensated by taking the winding or the number of the strings to be of order $N$. Let us suppose as above that $nM=N$ and we take the state \eqref{eq:conical defect state in symmetric product}.
In thermal $\text{AdS}_3$, the (holomorphic) ground state energy arises as follows:
\be 
-\frac{cN}{24}+\frac{cn(M^2-1)}{24M}=-\frac{cN}{24M^2}\ ,
\ee
where the first and second term correspond to the sphere and the torus contribution respectively. In the conical defect, the ground state energy arises directly from the sphere diagram, see eq.~\eqref{eq:conical defect sphere contribution}.
Interpreting the sphere contribution as the classical part of the bulk theory, we see that the conical defect is classically dual to the $M$-twisted sector ground states, but to see the $M$-twisted sector from thermal $\text{AdS}_3$, strong quantum effects are necessary.

This provides us with another important lesson. We should think of different backgrounds as different \emph{duality frames} of the same underlying theory. Different sectors of the theory are easier to access from one perspective or another. In particular, the classical description around these backgrounds captures different sectors of the symmetric product orbifold.

\section{Handlebodies and wormholes} \label{sec:higher genus}
In this section, we will comment about more complicated geometries, the implications for the AdS/CFT correspondence and stringy black hole physics. 

In general, we would like to consider $n$ Riemann surfaces $\Sigma_{g_i}$ (with fixed spin structures). From a gravity point of view, the partition function on these disconnected spaces can be computed by summing over all possible three-manifolds with the given boundary and compatible spin structures,\footnote{Conceivably, this sum should also extend over orbifold geometries such as the conical defect that we discussed above. There is considerable debate about this issue \cite{Benjamin:2020mfz, Cotler:2020ugk, Maxfield:2020ale}.}
\be 
\mathfrak{Z}_{\Sigma_{g_1} \sqcup \cdots\sqcup\, \Sigma_{g_n}}^{\text{Sym}}=\sum_{\genfrac{}{}{0pt}{}{\mathcal{M}_3\text{ spin}}{\partial \mathcal{M}_3=\Sigma_{g_1} \sqcup \cdots \sqcup\,\Sigma_{g_n}}} \mathfrak{Z}_{\mathcal{M}_3}\ , \label{eq:geometry sum}
\ee
where $\mathfrak{Z}_{\mathcal{M}_3}$ is the partition function evaluated on the spin manifold $\mathcal{M}_3$.\footnote{Every orientable 3-manifold is spin, see e.g.~\cite[page 46]{Kirby}.
}
In order for the geometries to correspond to saddles of Euclidean gravity, one should also impose that they are locally $\mathrm{AdS}_3$, i.e.~hyperbolic manifolds. However, also non-saddles can contribute which was demonstrated recently in the case of 3d gravity for the topology $\mathcal{M}_3=\mathbb{T}^2 \times [0,1]$ \cite{Cotler:2020ugk}.\footnote{Historically, it was often assumed that the putative dual CFT to pure 3d gravity is extremal and holomorphically factorized. In this holomorphic setting and for other holomorphic quantities like the elliptic genus, the sum over bulk geometries seems to work well \cite{Dijkgraaf:2000fq, Witten:2007kt, Manschot:2007zb,  Manschot:2007ha}.}
We shall see that the corresponding formula looks very different for the tensionless limit of string theory.

Notice that we work in the grand canonical ensemble with fixed chemical potential. From the boundary CFT point of view, the grand canonical partition function naturally factorizes on disconnected boundary components,
\be 
\mathfrak{Z}_{\Sigma_{g_1} \sqcup \cdots\sqcup\, \Sigma_{g_n}}^{\text{Sym}}=\prod_{i=1}^n\mathfrak{Z}_{\Sigma_{g_i}}\ .
\ee
Interpreted in terms of the canonical partition function, we would hence consider the collection of symmetric product orbifolds $\mathrm{Sym}^{N_i}(\mathbb{T}^4)$ where only $N_1+\cdots +N_n=N$ is kept fixed. One could refine the analysis by introducing different chemical potentials for the different boundary components.

Unfortunately, there are major technical difficulties that prevented us to compute the string partition function beyond the cases that we have considered:
\begin{enumerate}
\item The thermal $\text{AdS}_3$/BTZ/conical defect geometry can be obtained by an orbifold with abelian orbifold group that just acts by diagonal matrices in $\mathrm{SL}(2,\mathbb{C})$, see eqs.~\eqref{eq:thermal AdS3 orbifold action} and \eqref{eq:conical defect orbifold action}. Consequently it is quite simple to implement this orbifold on the worldsheet partition function. The third orbifold action that leads to a torus boundary in Section~\ref{subsec:single torus boundary} could also be computable. However, the orbifold generators are non-diagonalizable, which means that one should work in a coherent state basis on the worldsheet. We have not tried to do this.
For any other boundary geometry, the corresponding orbifold group is vastly more complicated (a non-elementary Kleinian group). For example the genus 2 handlebody can be obtained by orbifolding with a Schottky group that is isomorphic to a free group on two free generators.
\item All `interesting' geometries beyond those we have considered have at least genus 2 boundaries. This is explained in Section~\ref{subsec:multiple torus boundaries}. For a genus 2 boundary, we actually expect the worldsheet torus partition function to \emph{vanish}, since there are no branched covering maps $\Gamma: \Sigma_1 \to \Sigma_2$. Thus, we would actually need to compute also the genus $g \ge 2$ worldsheet partition function to learn anything interesting.
\end{enumerate}
In view of these difficulties, we will not attempt to compute these partition functions explicitly, but instead assume that they behave similarly to the cases we could compute. In particular, we expect the following general statement to always hold true:\footnote{This statement is expected to have a generalization to correlation functions as well, although we do not need it here. This is discussed in \cite{Eberhardt:2019ywk, Eberhardt:2020akk} in the case of global $\text{AdS}_3$, but we expect that the analogous statement for general bulk manifolds is also true.}
\begin{verse} 
\emph{The tensionless string partition function receives only contributions from \\ worldsheets $\Sigma_g$ for $g \ge 1$ that cover the boundary of $\mathcal{M}_3$ holomorphically.}
\end{verse}
Evidence for this proposal was collected in a number of circumstances: \cite{Eberhardt:2019ywk} discussed the case where both the worldsheet and the boundary is a sphere.\footnote{In order to get non-trivial covering maps, one has to look at correlators that introduce punctures in the two spheres. The relevant correlators were also studied using a different method in \cite{Hikida:2020kil}.} This was subsequently generalized to higher genus worldsheets on global $\mathcal{M}_3=\text{AdS}_3$, i.e.~$\partial \mathcal{M}_3=\mathbb{CP}^1$. Finally, we have established that the same is true in the case of a genus 1 worldsheet in thermal $\text{AdS}_3$ in Section~\ref{sec:thermal AdS3 partition function} and for the conical defect geometry in Section~\ref{sec:conical defect}.\footnote{The localization for thermal $\text{AdS}_3$ was already anticipated in \cite{Eberhardt:2018ouy}.} We shall assume in the following discussion that this paradigm holds true in full generality. Notice in particular that for a disconnected boundary $\partial \mathcal{M}_3 =\Sigma_{g_1} \sqcup \cdots \sqcup\, \Sigma_{g_n}$, this means that the worldsheet covers \emph{either} of the boundary components.
While partial proofs and evidence of this statement is available, it would be very important to give a general argument for the validity of this statement, since it lies at the heart of the $\text{AdS}_3/\text{CFT}_2$ correspondence.

There is one notable exception that we excluded from the statement. The sphere does not have any moduli and the volume of its automorphism group is infinite. Thus, it is much more subtle than the higher genus terms. There is never a holomorphic map from a spherical worldsheet to the boundary. The localization statement hence seems to suggest a vanishing answer which we saw explicitly in the case of thermal $\text{AdS}_3$ and the conical defects is not true. Instead the sphere contribution tends to shift the chemical potential, see eqs.~\eqref{eq:sphere sigma replacement} and \eqref{eq:sphere sigma replacement conical defect}.\footnote{The only case, where such a holomorphic map exists is the case of global $\text{AdS}_3$, where the boundary is itself a sphere. The sphere partition function of the boundary CFT is divergent and the finite piece can be removed by counterterms, which indicates that a version of localization still holds for the sphere.}
\subsection{Higher genus boundary}
Let us first discuss a single higher genus boundary $\Sigma_g$. The general expectation is that the worldsheet theory localizes on all holomorphic covering maps $\Gamma:\Sigma_{g'}' \mapsto \Sigma_g$. The existence of such covering maps can be understood fully algebraically, which is the subject of Hurwitz theory.

In the following we will only need the unramified case. The only topological constraint for such a map to exist follows by comparing Euler characteristics,\footnote{It is sufficient for a covering map to exist topologically, the Riemann existence theorem guarantees that in this case it will also exist holomorphically.}
\be 
2-2g'=\chi(\Sigma_{g'}')=d \chi(\Sigma_{g})=d(2-2g)\ , \label{eq:Euler characteristics}
\ee
where $d \in \mathbb{Z}_{\ge 1}$ is the degree of the covering map. Thus, the genus 0, 1 and $\ge 2$ partition functions behave very differently. 
\begin{enumerate}
\item $g=0$. In this case, the only contribution to the partition function arises from a spherical worldsheet (or a collection of disconnected spherical worldsheets) that cover the boundary of global $\text{AdS}_3$ once.
\item $g=1$. This is the case we have considered so far in this paper. In this case we necessarily also have $g'=0$, i.e.~only worldsheet tori contribute to the torus partition function.\footnote{Strictly speaking, also the worldsheet sphere contributes, but since the sphere has no moduli, we view this as an exceptional case.} However, the degree of the covering map is unconstrained in this case. This explains in hindsight, why we did not need to include the higher genus worldsheets to fully account for the torus partition function and why our result turned out to be exact.
\item $g \ge 2$. In this case we have $g' \ge g$, depending on the degree of the map. $g'$ is otherwise unconstrained.
\end{enumerate}
In the following we continue to discuss the $g \ge 2$ case. There are a priori many bulk manifolds that could contribute to the CFT partition function. Particularly simple hyperbolic manifolds are handlebody geometries. 
However, starting from genus 2, there are also non-handlebody bulk geometries, see also the discussion in Section~\ref{subsec:single higher genus boundary}. It is also possible that one can include various orbifold geometries as saddlepoints of the gravitational path integral.

We expect that the string picture is much simpler. By computing the string one-loop partition function, we have given evidence so far that string propagation on thermal $\text{AdS}_3$, on the various black hole geometries in $\text{AdS}_3$ that are related to thermal $\text{AdS}_3$ by a boundary $\mathrm{SL}(2,\mathbb{Z})$ transformation (modulo the choice of spin structure), and on the conical defect geometry is \emph{equivalent}.  We believe that the situation is qualitatively similar for a higher genus boundary. In other words, we conjecture that string theory on the various handlebody and non-handlebody geometries with the same higher genus surface (and same choice of boundary spin structure) is equivalent. The vacuum of one string background can be reinterpreted in another background as a highly excited string state. There should be various phase transitions in the moduli space $\mathcal{M}_{g,0}$ because different configurations dominate the partition function.

It is nothing new that string theory is insensitive to topological features of the target manifold, but this example takes this phenomenon to the extreme. Thus, in this string theory example, one can choose \emph{any} geometry compatible with the boundary spin structure:
\be 
\mathfrak{Z}_{\Sigma_g}^{\text{Sym}}=  \mathfrak{Z}_{\mathcal{M}_3}\ .
\ee

\paragraph{$N$-dependence.} There is a confusing aspect about this proposal that again has to do with the grand canonical ensemble. Since $g_\text{string}^2 \sim N^{-1}$, we would expect that the genus $g$ contribution to the partition function is suppressed. Actually, all higher genus surfaces contribute at the same order in string theory, but only finitely many of them do for fixed $N$. This is in agreement with the higher genus partition function of the symmetric orbifold, where higher genus surfaces are not suppressed in $N^{-1}$ \cite{Bantay:1998fy, Haehl:2014yla}.
To understand this, consider eq.~\eqref{eq:Euler characteristics}. A genus $g'$ worldsheet will contribute $d$ units of fundamental strings and contributes a prefactor $g_\text{string}^{2g'-2}=N^{1-g'}=N^{d(1-g)}$. Here we made use of the relation \eqref{eq:Euler characteristics}. Accounting for disconnected worldsheets, the canonical partition function receives a prefactor $N^{N(1-g)}$. This factor can be cancelled in the boundary theory by including the counterterm $\alpha \int_{\Sigma_g} \sqrt{g} R$ in the action of the boundary theory for appropriate $\alpha$. We note that this problem does not arise in thermal $\text{AdS}_3$. It would be good to understand this better.\footnote{A perhaps more natural interpretation is the following. The string worldsheet for the covering surfaces is essentially confined to a two-dimensional subspace of $\text{AdS}_3$, which could change the effective string coupling constant of these worldsheets to be $N$-independent. 
This would be similar to the mechanism explained in \cite{Verlinde:2002ig, Giveon:2005mi}. 
We leave a precise understanding of this for future work.}

\subsection{Disconnected boundaries}
The sum over geometries \eqref{eq:geometry sum} implies a serious paradox for the AdS/CFT correspondence. While the left-hand-side of the equation is factorized, the factorization is broken on the right-hand-side by the existence of wormhole geometries that connect the various boundary components.

In gravity theories, this is usually interpreted that the dual theory is an ensemble of theories. The ensemble average breaks the factorization on the left-hand-side of \eqref{eq:geometry sum}. This has been explored in detail in JT gravity in \cite{Saad:2019lba, Stanford:2019vob} and there are signs that a similar statement also holds in pure 3d gravity \cite{Cotler:2020ugk}.

That resolution is not possible in the present example, because there is a single boundary CFT. Consider for example a saddle of the supergravity action of the form $\Sigma_2 \times [0,1]$, where $\Sigma_2$ is a genus 2 surfaces, see the discussion in Section~\ref{subsec:two higher genus boundaries}.
There are also disconnected saddles with the same boundaries, for example the disjoint union of two genus 2 handlebodies. In analogy to the indistinguishability of thermal $\text{AdS}_3$ and the conical defect geometry (or any of the $\mathrm{SL}(2,\mathbb{Z})$ family of black holes with compatible spin structure), we conjecture that these geometries are indistinguishable for the tensionless string. In general, we conjecture the following property of the tensionless string:
\begin{verse}
\emph{The tensionless string distinguishes hyperbolic geometries (i.e.~saddles of\\  the gravity action) only according to the boundary geometry.}
\end{verse}
Assuming this conjecture to hold true in general, we have to suitably modify \eqref{eq:geometry sum} for the disconnected case. We can always pick a disconnected manifold $\mathcal{M}_3=\mathcal{M}_{3,1} \sqcup \cdots \sqcup \, \mathcal{M}_{3,n}$, where the spin structure of $\mathcal{M}_{3,i}$ is compatible with the spin structure of $\Sigma_{g_i}$. Thus we have for the tensionless string
\be 
\mathfrak{Z}_{\Sigma_{g_1} \sqcup \cdots\sqcup\, \Sigma_{g_n}}^{\text{Sym}}=\mathfrak{Z}_{\mathcal{M}_3}=\prod_{i=1}^n \mathfrak{Z}_{\mathcal{M}_{3,i}}\ , \label{eq:tensionless string manifold sum}
\ee
and thus also the bulk part of \eqref{eq:tensionless string manifold sum} factorizes.

We emphasize that the identification is not trivial. While the factorization becomes obvious when we choose a disconnected representation of the partition function, it becomes obscure when we choose a connected representation. The entire spectrum should rearrange itself as we discussed it for the conical defect in Section~\ref{subsec:conical defect rearrangement}. In particular, from the point of view of the connected geometry, large quantum corrections are necessary to the classical action to achieve a factorized partition function. It would be interesting to understand the rearrangement in more detail in simple examples. 

We finally mention that one can generalize the grand canonical ensemble for several boundaries to have independent chemical potentials for the different boundaries, which corresponds to fixing the field strength of the $B$-field to have different values close to the different boundaries. We can then refine \eqref{eq:tensionless string manifold sum} as follows:
\be 
\mathfrak{Z}_{\Sigma_{g_1} \sqcup \cdots\sqcup\, \Sigma_{g_n}}^{\text{Sym}}(t,z,\sigma_1,\dots,\sigma_n)=\mathfrak{Z}_{\mathcal{M}_3}(t,z,\sigma_1,\dots,\sigma_n)=\prod_{i=1}^n \mathfrak{Z}_{\mathcal{M}_{3,i}}(t,z,\sigma_i)\ . \label{eq:tensionless string manifold sum refined}
\ee

One might also view the grand canonical ensemble as a kind of ensemble average. We emphasize however that the grand canonical ensemble and the collection of canonical ensembles contain the same information and one could translate our analysis to the canonical ensemble, where not such `average' occurs.
\section{Discussion}\label{sec:discussion}
\subsection{Summary}
In this paper, we have studied the tensionless string on  hyperbolic manifolds with various Riemann surfaces as boundaries. Because of computational simplicity, we mainly focused on the genus 1 case, where we analyzed thermal $\text{AdS}_3$, the BTZ black hole and the conical defect geometry. We learned and confirmed some physical lessons. They are mostly reflections and consequences of each other. We have provided evidence for the following main statements.
\begin{enumerate}
\item Tensionless string theory on different hyperbolic manifolds with the same boundary surfaces and the same boundary spin structure is \emph{equivalent}.
\item This equivalence should be understood as a duality. Some states arise in a simple (classical) manner on some manifolds, while they correspond to highly excited states on others. In particular, the string sphere contribution in one background can become a higher genus contribution in another background.
\item These dualities solve the factorization problem that is present for example in JT gravity \cite{Teitelboim:1983ux, Jackiw:1984je, Saad:2019lba}  in this specific setting. The disconnected and connected contributions to a wormhole partition function agree and thus the partition function manifestly factorizes also in the bulk.
\item The theory does not contain any black holes. Instead, black holes become a single very long string. This is an explicit realization of the black hole/string transition \cite{Susskind:1993ws, Horowitz:1996nw}.
\item On a more technical level, we found it convenient to compute string partition functions in the grand canonical ensemble. We amassed further evidence for the localization of worldsheet correlators to covering spaces.
\end{enumerate}
\subsection{Discussion} \label{subsec:discussion}
There are several points that are not well understood that we would like to mention.
\paragraph{Grand canonical ensemble.} Our use of the grand canonical ensemble in string perturbation theory was ad hoc. To make the computations presented in this article more rigorous and to extend them to more complicated situations, it would be important to develop the grand canonical formulation better. In particular, it would be very illuminating if such a formalism could be derived from string field theory.
We should also note that the $g_\text{string}$ dependence in this situation is somewhat confusing. In particular we have seen that different higher genus surfaces can contribute at the same order, but the perturbative series still makes sense since it truncates for given $N$ at sufficiently high order (and is hence exact).
\paragraph{Sphere diagram.} We fixed the contribution of the sphere diagram to the string partition function using the constraints of conformal symmetry in the boundary and from extrapolating the supergravity result to $k=1$. It would be much more satisfying to give a first principle derivation of the value within string perturbation theory. This requires either a regularization of the usual string perturbation theory prescription or a different gauge fixing of the $\mathrm{SL}(2,\mathbb{C})$ automorphism group of the Riemann sphere as is done in \cite{Erbin:2019uiz}, see also \cite{Liu:1987nz, Tseytlin:1988tv}.
\paragraph{Backreaction.} We started with the `near-horizon limit' of one NS5-brane, which has mass of order $1/g_\text{string}^2 \sim N$. However, we put $N$ fundamental strings in the background; a single string winding $N$ times around the boundary of $\text{AdS}_3$ also has mass of order $N$ and thus its backreaction on the geometry is no longer negligible. This is exactly what we found: the $N$-fold winding string gives a description of the BTZ black hole. However it is actually questionable why we just trust string perturbation theory in this regime. Given that the partition functions we computed match exactly those of the symmetric orbifold, this seems to be the correct description. However, it would be interesting to understand this point better.
\paragraph{Non-perturbative effects.} The discussion in this paper was limited to the perturbative sector of string theory, but in general one also expects D-instanton contributions to be present. In the symmetric product orbifold, there is no indication of non-perturbative effects and one hence might expect that the D-instanton corrections are absent. As a first step in this direction one should classify possible worldsheet boundary states.
It would be good to understand this more directly analogously to the recent advances in another potentially provable version of holography -- the duality between $c=1$ strings and Matrix Quantum Mechanics \cite{Balthazar:2019rnh, Balthazar:2019ypi, Sen:2019qqg, Sen:2020cef}.\footnote{Also in the context of the $c=1$ string, only certain ZZ-instantons were needed on the worldsheet in order to reproduce the answer of Matrix Quantum Mechanics. }
\paragraph{Non-saddle contributions.} In 3d gravity, one should potentially also include geometries that are not saddles, such as the Seifert manifolds considered in \cite{Maxfield:2020ale}. Since these are not valid string backgrounds, we do not know how to set up string perturbation theory around these backgrounds and we cannot exclude that we should include these backgrounds in the full string path integral. Given that the other saddles are already included in the perturbative partition function around a given saddle, we find it reasonable that the string partition function that we have computed also accounts for these non-saddle contributions.
\subsection{Future directions}
Let us outline a couple of interesting questions for future research.
\paragraph{Euclidean wormhole.} While we have only computed the partition function on geometries with a single torus boundary, it would be important to confirm that the picture that we developed indeed holds true. While this is probably not possible in closed form, one could try to evaluate the partition function somewhat more abstractly. The general expectation is that one can write it as a sum over $\mathbb{T}^4$ partition functions evaluated on covering surfaces of the boundary.
\paragraph{Proof of localization.} We have made use of the localization result that the worldsheet theory localizes to covering spaces. While for thermal AdS$_3$, we can directly check the validity of this assertion, we had to assume it in the higher genus case.
An important technical open question is to give a full proof of this property. For correlators with a genus 0 boundary, this proposal passes strong consistency checks, such as the Ward identities of the current algebra on the worldsheet \cite{Eberhardt:2019ywk, Eberhardt:2020akk}. These arguments are however rather involved and do not constitute a full proof or a conceptually clear understanding. It might be technically easier to show localization in a topologically twisted version of the correspondence \cite{Li:2020nei, Li:2020zwo}.
\paragraph{Other instances.} Our computations give hints about how string theory manages to cure various problems that are present in quantum theories of gravity without known UV completion, such as JT gravity or possibly also pure Einstein gravity on $\text{AdS}_3$, whose existence is unclear. At least in this specific instance various contributions to the string path integral conspired to achieve factorization of the partition function. The ingredients are however quite specific to the considered setting and holomorphic coverings etc.~are expected to be replaced by different concepts in more complicated theories.

There are related $\text{AdS}_3/\text{CFT}_2$ examples like $\text{AdS}_3 \times \text{S}^3 \times \text{S}^3 \times \text{S}^1$ that seem to exhibit very similar properties. There are two fluxes $k^+$ and $k^-$ through the two 3-spheres and varying them gives a one-parameter generalization of the story we have explored in this paper. This background was discussed intensely in the literature and while for $k^+=1$ there is a convincing proposed dual CFT, the symmetric orbifold of $\mathrm{S}^3 \times \mathrm{S}^1$, no such explicit proposal exists for general fluxes \cite{Gukov:2004ym, Tong:2014yna, Eberhardt:2017pty, Eberhardt:2019niq}. It would be worthwhile to understand these features also in this background.

One particularly important and perhaps most-studied instance of AdS/CFT is the correspondence between $\text{AdS}_5 \times \text{S}^5$ and $\mathcal{N}=4$ SYM \cite{Maldacena:1997re, Witten:1998qj}. In this case, no ensemble average should or can be taken and thus one also needs to resolve the factorization problem. There exist (stable) wormhole solutions in this context \cite{Maldacena:2004rf}. It is tempting to think that a similar picture as the one proposed in the present context applies. From a dual CFT point of view, the study of tensionless strings was pioneered in \cite{Sundborg:1999ue, Sundborg:2000wp}.
 In $\text{AdS}_5 \times \text{S}^5$, one can continuously vary the radius of $\text{AdS}_5$ and thus better understand how this picture connects to usual supergravity computations. 

\paragraph{Deformations.}
For $\text{AdS}_3\times \text{S}^3 \times \mathbb{T}^4$, the jump to $k=2$ is rather drastic and the features of the model change completely.\footnote{In the Euclidean setting, some aspects of our analysis might continue to hold for the long string sector of the $k\ge 2$ background, that is conjectured to be captured by a symmetric product orbifold of an $\mathcal{N}=4$ Liouville theory \cite{Seiberg:1999xz, Eberhardt:2019qcl, Dei:2019osr}. The theory exhibits a continuum which makes the computation of the partition function much more subtle.} The continuum signals a singularity of this point in the moduli space \cite{Seiberg:1999xz}. 
Instead, one could try to deform the theory away from the (non-singular) symmetric product orbifold by turning on Ramond-Ramond flux on $\mathbb{T}^4$. In the CFT, this can be done in conformal perturbation theory and one could try to understand how the spectrum gets deformed in the various backgrounds at least to leading order in conformal perturbation theory, see e.g.~\cite{Carson:2014ena}. One expects from U-duality that this deformation brings one closer to the supergravity point of the system \cite{Larsen:1999uk, David:2002wn}.

\paragraph{Lorentzian picture.} Our consideration was entirely Euclidean, but the correspondence between strings on $\text{AdS}_3 \times \text{S}^3 \times \mathbb{T}^4$ at $k=1$ and the symmetric product orbifold is expected to hold also true in a Lorentzian spacetime. It would be important to work out the corresponding Lorentzian picture of our analysis. The spacetimes we have considered admit different analytic continuations \cite{Skenderis:2009ju}. In this context, similar ideas than the one presented in this paper were recently advocated by Jafferis \cite{Jafferis_talk}.

\paragraph{Symmetries of the unhiggsed phase of string theory.} It is tantalizing to speculate that most of what we have said in this paper is a consequence of the enormous underlying symmetry principle of unhiggsed string theory that allows only for a unique answer of the string partition function. At present, there is insufficient understanding of these symmetries to answer this question. At the level of the chiral algebra of the symmetric orbifold, the corresponding symmetry is the so-called Higher Spin Square, which is a very big extension of the higher spin symmetry \cite{Gaberdiel:2015mra, Gaberdiel:2015wpo}. 

\section*{Acknowledgements}
I would like to thank Andrea Dei, Daniel Kapec, Raghu Mahajan, Juan Maldacena, Shiraz Minwalla, Mukund Rangamani for useful conversations. I thank Matthias Gaberdiel and Rajesh Gopakumar for discussions and collaboration on related topics. I especially thank Shota Komatsu for many stimulating discussions and Edward Witten, whose questions have initiated this project. I thank Andrea Dei, Matthias Gaberdiel and Shota Komatsu for comments on a preliminary version of the manuscript.
I thank the della Pietra family for support at the IAS. 
\appendix

\section{The bosonic string on thermal \texorpdfstring{$\text{AdS}_3$}{AdS3}} \label{app:bosonic}
In this appendix, we show how to reproduce the known thermal bosonic $\text{AdS}_3$ partition function by taking a $\mathbb{Z}$-orbifold of the $\mathfrak{sl}(2,\mathbb{R})_k$ model, or rather of the Euclidean analogue -- the $\mathbb{H}^3$-model. This serves as a testing ground for the technology we employ in this paper.
\subsection{The partition function of the \texorpdfstring{$\mathbb{H}^3$}{H3}  model}
We start by recalling the partition function of the $\mathbb{H}^3$-model describing global Euclidean $\text{AdS}_3$. It is derived in \cite{Maldacena:2000hw}. Let us introduce a chemical potential $\theta$ for $\mathfrak{sl}(2,\mathbb{C})$. We normalize the charge such that the adjoint representation has charges $1$, $0$ and $-1$. The partition function reads \cite{Gawedzki:1991yu, Maldacena:2000hw}\footnote{The constant prefactor $\tfrac{1}{2} \sqrt{k-2}$ is somewhat subtle to fix, since it depends on the measure that one chooses for the $\mathfrak{sl}(2,\mathbb{C})$ spin. Choosing the Lebesgue measure results in this constant.}
\be 
Z_{\mathbb{H}^3_k}(\theta;\tau)=\frac{\sqrt{k-2}\exp\left(-\frac{(k-2)\pi (\Im \theta)^2}{\Im \tau}\right)}{2\sqrt{\Im \tau} \abs{\vartheta_1(\theta;\tau)}^2}\ .
\ee
The theta-function in the denominator accounts for the oscillator contribution to the partition function, whereas the remaining part comes from the integration over all possible $\mathfrak{sl}(2,\mathbb{C})$ spins. The partition function is modular covariant, as is appropriate for a non-holomorphic Jacobi form of index $(-\frac{k}{4},-\frac{k}{4})$:
\be 
Z_{\mathbb{H}^3_k} \left(\frac{\theta}{c\tau+d}; \frac{a \tau+b}{c \tau+d} \right)= \abs{\exp\left(-\frac{\pi i k \theta^2 c}{2(c \tau+d)} \right)}^2 Z_{\mathbb{H}^3_k}(\theta;\tau)\ . 
\ee
\subsection{The partition function on thermal \texorpdfstring{$\text{AdS}_3$}{AdS3}} \label{subapp:bosonic orbifold}
We now perform the orbifold to obtain the worldsheet theory that describes the boundary torus. By definition, states with charge $Q$ under $\mathfrak{sl}(2,\mathbb{C})$ receive a factor $\mathrm{e}^{2\pi i Q t}$ under the orbifold action. Thus,
\be 
\text{\scalebox{.7}{$m$}}\underset{0}{\raisebox{-5pt}{\text{\scalebox{2}{$\square$}}}}= Z_{\mathbb{H}^3_k}(mt;\tau)\ .
\ee
Under modular transformations
\be 
\text{\scalebox{.7}{$m$}}\underset{n}{\raisebox{-5pt}{\text{\scalebox{2}{$\square$}}}} \xrightarrow{\begin{pmatrix}
a & b \\ c & d
\end{pmatrix}}
\text{\scalebox{.7}{$am+b n$}}\underset{cm+dn}{\raisebox{-5pt}{\text{\scalebox{2}{$\square$}}}} \ . \label{eq:orbifold transformation}
\ee
Thus,
\begin{align}
\text{\scalebox{.7}{$am$}}\underset{cm}{\raisebox{-5pt}{\text{\scalebox{2}{$\square$}}}}&=Z_{\mathbb{H}^3_k} \left(mt; \frac{d \tau-b}{-c\tau+a}\right) \\
&=Z_{\mathbb{H}^3_k} \left(\frac{mt(-c\tau+a)}{-c\tau+a};\frac{d \tau-b}{-c\tau+a}\right) \\
&=Z_{\mathbb{H}^3_k} \left(mt(-c\tau+a);\tau\right) \left| \exp\left(\frac{\pi i k m^2 t^2(-c\tau+a)^2 c}{2(-c\tau+a)} \right)\right|^2 \ .
\end{align}
As anticipated, this only depends on the combinations $ma$ and $cm$. This fixes unambiguously
\be 
\text{\scalebox{.7}{$m$}}\underset{n}{\raisebox{-5pt}{\text{\scalebox{2}{$\square$}}}} =Z_{\mathbb{H}^3_k} (t(m-n\tau);\tau) \left| \exp\left(\frac{\pi i k t^2 n(m-n\tau)}{2}\right)\right|^2\ .
\ee
By inserting the definition, we can rewrite the result as
\be 
\text{\scalebox{.7}{$m$}}\underset{n}{\raisebox{-5pt}{\text{\scalebox{2}{$\square$}}}}=\frac{\sqrt{k-2}\, \mathrm{e}^{\frac{2\pi (\Im t_{m,n})^2}{\Im \tau}-\frac{k \pi(\Im t)^2}{\Im \tau}|m-n \tau|^2}}{2\sqrt{\Im \tau} \abs{\vartheta_1(t_{m,n};\tau)}^2}\ .
\ee
Here, $t_{m,n}=t(m-n\tau)$ such that $t_{1,0}=t$. The full partition function is then
\be 
Z_{\text{thermal $\text{AdS}_3$}}(t;\tau)=\mathcal{N} \sum_{m,n \in \mathbb{Z}} \text{\scalebox{.7}{$m$}}\underset{n}{\raisebox{-5pt}{\text{\scalebox{2}{$\square$}}}}\ .
\ee
The normalization factor $\mathcal{N}$ (that should correspond to the inverse volume of the group we are orbifolding by) can be determined as follows. The orbifold projection imposes a singlet condition on the charge. For a state of charge $(h,\bar{h})$ w.r.t.~left- and right-movers ($h$ and $\bar{h}$ are both real and not complex conjugates), we have
\be 
\mathcal{N}\sum_{m \in \mathbb{Z}} \mathrm{e}^{2\pi i m(h t-\bar{h}\bar{t})}=\mathcal{N}\delta_{\mathbb{Z}}(ht-\bar{h}\bar{t})\ .
\ee
Let us assume that $t$ is purely imaginary, which allows us to simplify further,
\be 
\mathcal{N}\delta_{\mathbb{Z}}(ht-\bar{h}\bar{t})=\mathcal{N} \delta_{\mathbb{Z}} (i \Im(t) (h+\bar{h}))=\frac{\mathcal{N}}{\Im(t)} \delta (h+\bar{h})\ .
\ee
Here, we used that the argument is purely imaginary and the delta-function can only be satisfied for vanishing $\Delta=h+\bar{h}$. This is what one would expect, since a field with vanishing $\Delta$ is invariant under the scaling $x \to \mathrm{e}^{2\pi it}x$ for $t \in i \mathbb{R}$. 
Thus, we should have $\mathcal{N} = \Im(t)$, since then this sum corresponds indeed to the projection of charges to the invariant ones with $h+\bar{h}=0$. For $\Re(t) \ne 0$, the normalization should not change. The normalization $\mathcal{N}$ should be inversely proportional to the size of the orbifold group and hence proportional to the size of the target space. This size only depends on $\Im(t)$ and hence we postulate that in general $\mathcal{N}=\Im(t)$.

We have hence derived
\be 
Z_{\mathbb{H}_k/\mathbb{Z}}(t;\tau)=\frac{1}{2}\Im t \sqrt{k-2} \sum_{m,n \in \mathbb{Z}} \frac{\mathrm{e}^{\frac{2\pi (\Im t_{m,n})^2}{\Im \tau}-\frac{k \pi(\Im t)^2}{\Im \tau}|m-n \tau|^2}}{\sqrt{\Im \tau} \abs{\vartheta_1(t_{m,n};\tau)}^2}\ .
\ee
This agrees with equation (27) of \cite{Maldacena:2000kv}.\footnote{Our expression has an additional factor of $\tfrac{1}{2}$. This factor is compensated in the string partition function by the fact that the sectors $(m,n)$ and $(-m,-n)$ lead to the same contribution.}

\bibliographystyle{JHEP}
\bibliography{bib}

\providecommand{\href}[2]{#2}\begingroup\raggedright\begin{thebibliography}{100}

\bibitem{Maldacena:1997re}
J.~M. Maldacena, \emph{{The Large $N$ limit of superconformal field theories
  and supergravity}}, \href{https://doi.org/10.1023/A:1026654312961,
  10.4310/ATMP.1998.v2.n2.a1}{\emph{Int. J. Theor. Phys.} {\bfseries 38} (1999)
  1113} [\href{https://arxiv.org/abs/hep-th/9711200}{{\ttfamily
  hep-th/9711200}}].

\bibitem{Berkovits:2007rj}
N.~Berkovits and C.~Vafa, \emph{{Towards a Worldsheet Derivation of the
  Maldacena Conjecture}},
  \href{https://doi.org/10.1088/1126-6708/2008/03/031}{\emph{AIP Conf. Proc.}
  {\bfseries 1031} (2008) 21}
  [\href{https://arxiv.org/abs/0711.1799}{{\ttfamily 0711.1799}}].

\bibitem{Berkovits:2019ulm}
N.~Berkovits, \emph{{Sketching a Proof of the Maldacena Conjecture at Small
  Radius}}, \href{https://doi.org/10.1007/JHEP06(2019)111}{\emph{JHEP}
  {\bfseries 06} (2019) 111}
  [\href{https://arxiv.org/abs/1903.08264}{{\ttfamily 1903.08264}}].

\bibitem{Gross:1987ar}
D.~J. Gross and P.~F. Mende, \emph{{String Theory Beyond the Planck Scale}},
  \href{https://doi.org/10.1016/0550-3213(88)90390-2}{\emph{Nucl. Phys. B}
  {\bfseries 303} (1988) 407}.

\bibitem{Gaberdiel:2018rqv}
M.~R. Gaberdiel and R.~Gopakumar, \emph{{Tensionless string spectra on
  $\mathrm{AdS}_{3}$}},
  \href{https://doi.org/10.1007/JHEP05(2018)085}{\emph{JHEP} {\bfseries 05}
  (2018) 085} [\href{https://arxiv.org/abs/1803.04423}{{\ttfamily
  1803.04423}}].

\bibitem{Giribet:2018ada}
G.~Giribet, C.~Hull, M.~Kleban, M.~Porrati and E.~Rabinovici,
  \emph{{Superstrings on $\mathrm{AdS}_{3}$ at $k= 1$}},
  \href{https://doi.org/10.1007/JHEP08(2018)204}{\emph{JHEP} {\bfseries 08}
  (2018) 204} [\href{https://arxiv.org/abs/1803.04420}{{\ttfamily
  1803.04420}}].

\bibitem{Eberhardt:2018ouy}
L.~Eberhardt, M.~R. Gaberdiel and R.~Gopakumar, \emph{{The Worldsheet Dual of
  the Symmetric Product CFT}},
  \href{https://doi.org/10.1007/JHEP04(2019)103}{\emph{JHEP} {\bfseries 04}
  (2019) 103} [\href{https://arxiv.org/abs/1812.01007}{{\ttfamily
  1812.01007}}].

\bibitem{Eberhardt:2019ywk}
L.~Eberhardt, M.~R. Gaberdiel and R.~Gopakumar, \emph{{Deriving the
  $\mathrm{AdS}_{3}/\mathrm{CFT}_{2}$ correspondence}},
  \href{https://doi.org/10.1007/JHEP02(2020)136}{\emph{JHEP} {\bfseries 02}
  (2020) 136} [\href{https://arxiv.org/abs/1911.00378}{{\ttfamily
  1911.00378}}].

\bibitem{Eberhardt:2020akk}
L.~Eberhardt, \emph{{$\mathrm{AdS}_{3}/\mathrm{CFT}_{2}$ at higher genus}},
  \href{https://doi.org/10.1007/JHEP05(2020)150}{\emph{JHEP} {\bfseries 05}
  (2020) 150} [\href{https://arxiv.org/abs/2002.11729}{{\ttfamily
  2002.11729}}].

\bibitem{Hikida:2020kil}
Y.~Hikida and T.~Liu, \emph{{Correlation functions of symmetric orbifold from
  AdS$_3$ string theory}},  \href{https://arxiv.org/abs/2005.12511}{{\ttfamily
  2005.12511}}.

\bibitem{Giribet:2020mkc}
G.~Giribet, \emph{{On string theory on $\mathrm{AdS}_3\times {M}_7$ in the
  tensionless limit}},
  \href{https://doi.org/10.1142/S0218271820300050}{\emph{Int. J. Mod. Phys. D}
  {\bfseries 29} (2020) 2030005}
  [\href{https://arxiv.org/abs/2003.02868}{{\ttfamily 2003.02868}}].

\bibitem{Karlhede:1986wb}
A.~Karlhede and U.~Lindström, \emph{{The Classical Bosonic String in the Zero
  Tension Limit}},
  \href{https://doi.org/10.1088/0264-9381/3/4/002}{\emph{Class. Quant. Grav.}
  {\bfseries 3} (1986) L73}.

\bibitem{Isberg:1993av}
J.~Isberg, U.~Lindström, B.~Sundborg and G.~Theodoridis, \emph{{Classical and
  Quantized Tensionless Strings}},
  \href{https://doi.org/10.1016/0550-3213(94)90056-6}{\emph{Nucl. Phys. B}
  {\bfseries 411} (1994) 122}
  [\href{https://arxiv.org/abs/hep-th/9307108}{{\ttfamily hep-th/9307108}}].

\bibitem{Gustafsson:1994kr}
H.~Gustafsson, U.~Lindström, P.~Saltsidis, B.~Sundborg and R.~van Unge,
  \emph{{Hamiltonian BRST Quantization of the Conformal String}},
  \href{https://doi.org/10.1016/0550-3213(95)00051-S}{\emph{Nucl. Phys. B}
  {\bfseries 440} (1995) 495}
  [\href{https://arxiv.org/abs/hep-th/9410143}{{\ttfamily hep-th/9410143}}].

\bibitem{Lindstrom:2003mg}
U.~Lindström and M.~Zabzine, \emph{{Tensionless Strings, WZW Models at
  Critical Level and Massless Higher Spin Fields}},
  \href{https://doi.org/10.1016/j.physletb.2004.01.035}{\emph{Phys. Lett. B}
  {\bfseries 584} (2004) 178}
  [\href{https://arxiv.org/abs/hep-th/0305098}{{\ttfamily hep-th/0305098}}].

\bibitem{Bakas:2004jq}
I.~Bakas and C.~Sourdis, \emph{{On the Tensionless Limit of Gauged WZW
  Models}}, \href{https://doi.org/10.1088/1126-6708/2004/06/049}{\emph{JHEP}
  {\bfseries 06} (2004) 049}
  [\href{https://arxiv.org/abs/hep-th/0403165}{{\ttfamily hep-th/0403165}}].

\bibitem{Maloney:2007ud}
A.~Maloney and E.~Witten, \emph{{Quantum Gravity Partition Functions in Three
  Dimensions}}, \href{https://doi.org/10.1007/JHEP02(2010)029}{\emph{JHEP}
  {\bfseries 02} (2010) 029} [\href{https://arxiv.org/abs/0712.0155}{{\ttfamily
  0712.0155}}].

\bibitem{Yin:2007gv}
X.~Yin, \emph{{Partition Functions of Three-Dimensional Pure Gravity}},
  \href{https://doi.org/10.4310/CNTP.2008.v2.n2.a1}{\emph{Commun. Num. Theor.
  Phys.} {\bfseries 2} (2008) 285}
  [\href{https://arxiv.org/abs/0710.2129}{{\ttfamily 0710.2129}}].

\bibitem{Giombi:2008vd}
S.~Giombi, A.~Maloney and X.~Yin, \emph{{One-loop Partition Functions of 3D
  Gravity}}, \href{https://doi.org/10.1088/1126-6708/2008/08/007}{\emph{JHEP}
  {\bfseries 08} (2008) 007} [\href{https://arxiv.org/abs/0804.1773}{{\ttfamily
  0804.1773}}].

\bibitem{Keller:2014xba}
C.~A. Keller and A.~Maloney, \emph{{Poincare Series, 3D Gravity and CFT
  Spectroscopy}}, \href{https://doi.org/10.1007/JHEP02(2015)080}{\emph{JHEP}
  {\bfseries 02} (2015) 080} [\href{https://arxiv.org/abs/1407.6008}{{\ttfamily
  1407.6008}}].

\bibitem{Benjamin:2019stq}
N.~Benjamin, H.~Ooguri, S.-H. Shao and Y.~Wang, \emph{{Light-cone modular
  bootstrap and pure gravity}},
  \href{https://doi.org/10.1103/PhysRevD.100.066029}{\emph{Phys. Rev. D}
  {\bfseries 100} (2019) 066029}
  [\href{https://arxiv.org/abs/1906.04184}{{\ttfamily 1906.04184}}].

\bibitem{Alday:2019vdr}
L.~F. Alday and J.-B. Bae, \emph{{Rademacher Expansions and the Spectrum of 2D
  CFT}},  \href{https://arxiv.org/abs/2001.00022}{{\ttfamily 2001.00022}}.

\bibitem{Benjamin:2020mfz}
N.~Benjamin, S.~Collier and A.~Maloney, \emph{{Pure Gravity and Conical
  Defects}},  \href{https://arxiv.org/abs/2004.14428}{{\ttfamily 2004.14428}}.

\bibitem{Maxfield:2020ale}
H.~Maxfield and G.~J. Turiaci, \emph{{The Path Integral of 3D Gravity Near
  Extremality; Or, Jt Gravity with Defects as a Matrix Integral}},
  \href{https://arxiv.org/abs/2006.11317}{{\ttfamily 2006.11317}}.

\bibitem{Castro:2011zq}
A.~Castro, M.~R. Gaberdiel, T.~Hartman, A.~Maloney and R.~Volpato, \emph{{The
  Gravity Dual of the Ising Model}},
  \href{https://doi.org/10.1103/PhysRevD.85.024032}{\emph{Phys. Rev. D}
  {\bfseries 85} (2012) 024032}
  [\href{https://arxiv.org/abs/1111.1987}{{\ttfamily 1111.1987}}].

\bibitem{Jian:2019ubz}
C.-M. Jian, A.~W. Ludwig, Z.-X. Luo, H.-Y. Sun and Z.~Wang, \emph{{Establishing
  Strongly-Coupled 3D AdS Quantum Gravity with Ising Dual Using All-Genus
  Partition Functions}},  \href{https://arxiv.org/abs/1907.06656}{{\ttfamily
  1907.06656}}.

\bibitem{Hawking:1982dh}
S.~Hawking and D.~N. Page, \emph{{Thermodynamics of Black Holes in
  Anti-de~Sitter Space}},
  \href{https://doi.org/10.1007/BF01208266}{\emph{Commun. Math. Phys.}
  {\bfseries 87} (1983) 577}.

\bibitem{Keller:2011xi}
C.~A. Keller, \emph{{Phase Transitions in Symmetric Orbifold CFTs and
  Universality}}, \href{https://doi.org/10.1007/JHEP03(2011)114}{\emph{JHEP}
  {\bfseries 03} (2011) 114} [\href{https://arxiv.org/abs/1101.4937}{{\ttfamily
  1101.4937}}].

\bibitem{Saad:2019lba}
P.~Saad, S.~H. Shenker and D.~Stanford, \emph{{JT Gravity as a Matrix
  Integral}},  \href{https://arxiv.org/abs/1903.11115}{{\ttfamily 1903.11115}}.

\bibitem{Stanford:2019vob}
D.~Stanford and E.~Witten, \emph{{JT Gravity and the Ensembles of Random Matrix
  Theory}},  \href{https://arxiv.org/abs/1907.03363}{{\ttfamily 1907.03363}}.

\bibitem{Cotler:2020ugk}
J.~Cotler and K.~Jensen, \emph{{$\mathrm{AdS}_3$ gravity and random CFT}},
  \href{https://arxiv.org/abs/2006.08648}{{\ttfamily 2006.08648}}.

\bibitem{Belin:2020hea}
A.~Belin and J.~de~Boer, \emph{{Random Statistics of OPE Coefficients and
  Euclidean Wormholes}},  \href{https://arxiv.org/abs/2006.05499}{{\ttfamily
  2006.05499}}.

\bibitem{Afkhami-Jeddi:2020ezh}
N.~Afkhami-Jeddi, H.~Cohn, T.~Hartman and A.~Tajdini, \emph{{Free Partition
  Functions and an Averaged Holographic Duality}},
  \href{https://arxiv.org/abs/2006.04839}{{\ttfamily 2006.04839}}.

\bibitem{Maloney:2020nni}
A.~Maloney and E.~Witten, \emph{{Averaging over Narain Moduli Space}},
  \href{https://arxiv.org/abs/2006.04855}{{\ttfamily 2006.04855}}.

\bibitem{Altland:2020ccq}
A.~Altland and J.~Sonner, \emph{{Late Time Physics of Holographic Quantum
  Chaos}},  \href{https://arxiv.org/abs/2008.02271}{{\ttfamily 2008.02271}}.

\bibitem{Susskind:1993ws}
L.~Susskind, \emph{{Some Speculations About Black Hole Entropy in String
  Theory}},  \href{https://arxiv.org/abs/hep-th/9309145}{{\ttfamily
  hep-th/9309145}}.

\bibitem{Horowitz:1996nw}
G.~T. Horowitz and J.~Polchinski, \emph{{A Correspondence Principle for Black
  Holes and Strings}},
  \href{https://doi.org/10.1103/PhysRevD.55.6189}{\emph{Phys. Rev. D}
  {\bfseries 55} (1997) 6189}
  [\href{https://arxiv.org/abs/hep-th/9612146}{{\ttfamily hep-th/9612146}}].

\bibitem{Giveon:2005mi}
A.~Giveon, D.~Kutasov, E.~Rabinovici and A.~Sever, \emph{{Phases of Quantum
  Gravity in $\mathrm{AdS}_3$ and Linear Dilaton Backgrounds}},
  \href{https://doi.org/10.1016/j.nuclphysb.2005.04.015}{\emph{Nucl. Phys. B}
  {\bfseries 719} (2005) 3}
  [\href{https://arxiv.org/abs/hep-th/0503121}{{\ttfamily hep-th/0503121}}].

\bibitem{Pakman:2009zz}
A.~Pakman, L.~Rastelli and S.~S. Razamat, \emph{{Diagrams for Symmetric Product
  Orbifolds}}, \href{https://doi.org/10.1088/1126-6708/2009/10/034}{\emph{JHEP}
  {\bfseries 10} (2009) 034} [\href{https://arxiv.org/abs/0905.3448}{{\ttfamily
  0905.3448}}].

\bibitem{Maskit:2012}
B.~Maskit, \emph{{Kleinian groups}}, vol.~287. Springer Science \& Business
  Media, 2012.

\bibitem{Berkovits:1999im}
N.~Berkovits, C.~Vafa and E.~Witten, \emph{{Conformal field theory of ${\rm
  AdS}$ background with Ramond-Ramond flux}},
  \href{https://doi.org/10.1088/1126-6708/1999/03/018}{\emph{JHEP} {\bfseries
  03} (1999) 018} [\href{https://arxiv.org/abs/hep-th/9902098}{{\ttfamily
  hep-th/9902098}}].

\bibitem{Giveon:1998ns}
A.~Giveon, D.~Kutasov and N.~Seiberg, \emph{{Comments on string theory on ${\rm
  AdS}_3$}}, \href{https://doi.org/10.4310/ATMP.1998.v2.n4.a3}{\emph{Adv.
  Theor. Math. Phys.} {\bfseries 2} (1998) 733}
  [\href{https://arxiv.org/abs/hep-th/9806194}{{\ttfamily hep-th/9806194}}].

\bibitem{deBoer:1998gyt}
J.~de~Boer, H.~Ooguri, H.~Robins and J.~Tannenhauser, \emph{{String theory on
  ${\rm AdS}_3$}},
  \href{https://doi.org/10.1088/1126-6708/1998/12/026}{\emph{JHEP} {\bfseries
  12} (1998) 026} [\href{https://arxiv.org/abs/hep-th/9812046}{{\ttfamily
  hep-th/9812046}}].

\bibitem{Witten:1999xp}
E.~Witten and S.-T. Yau, \emph{{Connectedness of the Boundary in the AdS / CFT
  Correspondence}},  \href{https://arxiv.org/abs/hep-th/9910245}{{\ttfamily
  hep-th/9910245}}.

\bibitem{Krasnov:2000zq}
K.~Krasnov, \emph{{Holography and Riemann surfaces}},
  \href{https://doi.org/10.4310/ATMP.2000.v4.n4.a5}{\emph{Adv. Theor. Math.
  Phys.} {\bfseries 4} (2000) 929}
  [\href{https://arxiv.org/abs/hep-th/0005106}{{\ttfamily hep-th/0005106}}].

\bibitem{Takhtajan:2002cc}
L.~A. Takhtajan and L.-P. Teo, \emph{{Liouville Action and Weil-Petersson
  Metric on Deformation Spaces, Global Kleinian Reciprocity and Holography}},
  \href{https://doi.org/10.1007/s00220-003-0878-5}{\emph{Commun. Math. Phys.}
  {\bfseries 239} (2003) 183}
  [\href{https://arxiv.org/abs/math/0204318}{{\ttfamily math/0204318}}].

\bibitem{Yin:2007at}
X.~Yin, \emph{{On Non-Handlebody Instantons in 3D Gravity}},
  \href{https://doi.org/10.1088/1126-6708/2008/09/120}{\emph{JHEP} {\bfseries
  09} (2008) 120} [\href{https://arxiv.org/abs/0711.2803}{{\ttfamily
  0711.2803}}].

\bibitem{Bers:1960}
L.~Bers, \emph{Simultaneous uniformization}, {\emph{Bull. Amer. Math. Soc.}
  {\bfseries 66} (1960) 94}.

\bibitem{Gotz:2006qp}
G.~Gotz, T.~Quella and V.~Schomerus, \emph{{The WZNW model on ${\rm
  PSU}(1,1\vert 2)$}},
  \href{https://doi.org/10.1088/1126-6708/2007/03/003}{\emph{JHEP} {\bfseries
  03} (2007) 003} [\href{https://arxiv.org/abs/hep-th/0610070}{{\ttfamily
  hep-th/0610070}}].

\bibitem{Martinec:2001cf}
E.~J. Martinec and W.~McElgin, \emph{{String theory on AdS orbifolds}},
  \href{https://doi.org/10.1088/1126-6708/2002/04/029}{\emph{JHEP} {\bfseries
  04} (2002) 029} [\href{https://arxiv.org/abs/hep-th/0106171}{{\ttfamily
  hep-th/0106171}}].

\bibitem{Martinec:2002xq}
E.~J. Martinec and W.~McElgin, \emph{{Exciting AdS orbifolds}},
  \href{https://doi.org/10.1088/1126-6708/2002/10/050}{\emph{JHEP} {\bfseries
  10} (2002) 050} [\href{https://arxiv.org/abs/hep-th/0206175}{{\ttfamily
  hep-th/0206175}}].

\bibitem{Porrati:2015eha}
J.~Kim and M.~Porrati, \emph{{On the central charge of spacetime current
  algebras and correlators in string theory on $\mathrm{AdS}_{3}$}},
  \href{https://doi.org/10.1007/JHEP05(2015)076}{\emph{JHEP} {\bfseries 05}
  (2015) 076} [\href{https://arxiv.org/abs/1503.07186}{{\ttfamily
  1503.07186}}].

\bibitem{Kutasov:1999xu}
D.~Kutasov and N.~Seiberg, \emph{{More comments on string theory on ${\rm
  AdS}_3$}}, \href{https://doi.org/10.1088/1126-6708/1999/04/008}{\emph{JHEP}
  {\bfseries 04} (1999) 008}
  [\href{https://arxiv.org/abs/hep-th/9903219}{{\ttfamily hep-th/9903219}}].

\bibitem{Erbin:2019uiz}
H.~Erbin, J.~Maldacena and D.~Skliros, \emph{{Two-Point String Amplitudes}},
  \href{https://doi.org/10.1007/JHEP07(2019)139}{\emph{JHEP} {\bfseries 07}
  (2019) 139} [\href{https://arxiv.org/abs/1906.06051}{{\ttfamily
  1906.06051}}].

\bibitem{Maldacena:1998bw}
J.~M. Maldacena and A.~Strominger, \emph{{${\rm AdS}_3$ black holes and a
  stringy exclusion principle}},
  \href{https://doi.org/10.1088/1126-6708/1998/12/005}{\emph{JHEP} {\bfseries
  12} (1998) 005} [\href{https://arxiv.org/abs/hep-th/9804085}{{\ttfamily
  hep-th/9804085}}].

\bibitem{Brown:1986nw}
J.~Brown and M.~Henneaux, \emph{{Central Charges in the Canonical Realization
  of Asymptotic Symmetries: an Example from Three-Dimensional Gravity}},
  \href{https://doi.org/10.1007/BF01211590}{\emph{Commun. Math. Phys.}
  {\bfseries 104} (1986) 207}.

\bibitem{Eberhardt:2019qcl}
L.~Eberhardt and M.~R. Gaberdiel, \emph{{String theory on $\mathrm{AdS}_3$ and
  the symmetric orbifold of Liouville theory}},
  \href{https://doi.org/10.1016/j.nuclphysb.2019.114774}{\emph{Nucl. Phys. B}
  {\bfseries 948} (2019) 114774}
  [\href{https://arxiv.org/abs/1903.00421}{{\ttfamily 1903.00421}}].

\bibitem{Rangamani:2007fz}
M.~Rangamani and S.~F. Ross, \emph{{Winding tachyons in BTZ}},
  \href{https://doi.org/10.1103/PhysRevD.77.026010}{\emph{Phys. Rev. D}
  {\bfseries 77} (2008) 026010}
  [\href{https://arxiv.org/abs/0706.0663}{{\ttfamily 0706.0663}}].

\bibitem{Polchinski:1985zf}
J.~Polchinski, \emph{{Evaluation of the One Loop String Path Integral}},
  \href{https://doi.org/10.1007/BF01210791}{\emph{Commun. Math. Phys.}
  {\bfseries 104} (1986) 37}.

\bibitem{Maldacena:2000kv}
J.~M. Maldacena, H.~Ooguri and J.~Son, \emph{{Strings in ${\rm AdS}_3$ and
  ${\rm SL}(2,\mathds{R})$ WZW model. Part 2. Euclidean black hole}},
  \href{https://doi.org/10.1063/1.1377039}{\emph{J. Math. Phys.} {\bfseries 42}
  (2001) 2961} [\href{https://arxiv.org/abs/hep-th/0005183}{{\ttfamily
  hep-th/0005183}}].

\bibitem{Dijkgraaf:1996xw}
R.~Dijkgraaf, G.~W. Moore, E.~P. Verlinde and H.~L. Verlinde, \emph{{Elliptic
  Genera of Symmetric Products and Second Quantized Strings}},
  \href{https://doi.org/10.1007/s002200050087}{\emph{Commun. Math. Phys.}
  {\bfseries 185} (1997) 197}
  [\href{https://arxiv.org/abs/hep-th/9608096}{{\ttfamily hep-th/9608096}}].

\bibitem{Maldacena:1999bp}
J.~M. Maldacena, G.~W. Moore and A.~Strominger, \emph{{Counting BPS Black Holes
  in Toroidal Type II String Theory}},
  \href{https://arxiv.org/abs/hep-th/9903163}{{\ttfamily hep-th/9903163}}.

\bibitem{Hartman:2014oaa}
T.~Hartman, C.~A. Keller and B.~Stoica, \emph{{Universal Spectrum of 2D
  Conformal Field Theory in the Large C Limit}},
  \href{https://doi.org/10.1007/JHEP09(2014)118}{\emph{JHEP} {\bfseries 09}
  (2014) 118} [\href{https://arxiv.org/abs/1405.5137}{{\ttfamily 1405.5137}}].

\bibitem{Haehl:2014yla}
F.~M. Haehl and M.~Rangamani, \emph{{Permutation Orbifolds and Holography}},
  \href{https://doi.org/10.1007/JHEP03(2015)163}{\emph{JHEP} {\bfseries 03}
  (2015) 163} [\href{https://arxiv.org/abs/1412.2759}{{\ttfamily 1412.2759}}].

\bibitem{Holzhey:1991bx}
C.~F. Holzhey and F.~Wilczek, \emph{{Black Holes as Elementary Particles}},
  \href{https://doi.org/10.1016/0550-3213(92)90254-9}{\emph{Nucl. Phys. B}
  {\bfseries 380} (1992) 447}
  [\href{https://arxiv.org/abs/hep-th/9202014}{{\ttfamily hep-th/9202014}}].

\bibitem{Sen:1995in}
A.~Sen, \emph{{Extremal Black Holes and Elementary String States}},
  \href{https://doi.org/10.1142/S0217732395002234}{\emph{Mod. Phys. Lett. A}
  {\bfseries 10} (1995) 2081}
  [\href{https://arxiv.org/abs/hep-th/9504147}{{\ttfamily hep-th/9504147}}].

\bibitem{Witten:1991yr}
E.~Witten, \emph{{On String Theory and Black Holes}},
  \href{https://doi.org/10.1103/PhysRevD.44.314}{\emph{Phys. Rev. D} {\bfseries
  44} (1991) 314}.

\bibitem{Kutasov:2000jp}
D.~Kutasov and D.~Sahakyan, \emph{{Comments on the Thermodynamics of Little
  String Theory}},
  \href{https://doi.org/10.1088/1126-6708/2001/02/021}{\emph{JHEP} {\bfseries
  02} (2001) 021} [\href{https://arxiv.org/abs/hep-th/0012258}{{\ttfamily
  hep-th/0012258}}].

\bibitem{Nakayama:2005pk}
Y.~Nakayama, S.-J. Rey and Y.~Sugawara, \emph{{D-Brane Propagation in
  Two-Dimensional Black Hole Geometries}},
  \href{https://doi.org/10.1088/1126-6708/2005/09/020}{\emph{JHEP} {\bfseries
  09} (2005) 020} [\href{https://arxiv.org/abs/hep-th/0507040}{{\ttfamily
  hep-th/0507040}}].

\bibitem{Atick:1988si}
J.~J. Atick and E.~Witten, \emph{{The Hagedorn Transition and the Number of
  Degrees of Freedom of String Theory}},
  \href{https://doi.org/10.1016/0550-3213(88)90151-4}{\emph{Nucl. Phys. B}
  {\bfseries 310} (1988) 291}.

\bibitem{Berkooz:2007fe}
M.~Berkooz, Z.~Komargodski and D.~Reichmann, \emph{{Thermal $\mathrm{AdS}_3$,
  BTZ and Competing Winding Modes Condensation}},
  \href{https://doi.org/10.1088/1126-6708/2007/12/020}{\emph{JHEP} {\bfseries
  12} (2007) 020} [\href{https://arxiv.org/abs/0706.0610}{{\ttfamily
  0706.0610}}].

\bibitem{Lin:2007gi}
F.-L. Lin, T.~Matsuo and D.~Tomino, \emph{{Hagedorn Strings and Correspondence
  Principle in $\mathrm{AdS}_3$}},
  \href{https://doi.org/10.1088/1126-6708/2007/09/042}{\emph{JHEP} {\bfseries
  09} (2007) 042} [\href{https://arxiv.org/abs/0705.4514}{{\ttfamily
  0705.4514}}].

\bibitem{deLange:2018mri}
P.~De~Lange, A.~Maloney and E.~Verlinde, \emph{{Monstrous Product CFTs in the
  Grand Canonical Ensemble}},
  \href{https://arxiv.org/abs/1807.06200}{{\ttfamily 1807.06200}}.

\bibitem{Vafa:1986wx}
C.~Vafa, \emph{{Modular Invariance and Discrete Torsion on Orbifolds}},
  \href{https://doi.org/10.1016/0550-3213(86)90379-2}{\emph{Nucl. Phys. B}
  {\bfseries 273} (1986) 592}.

\bibitem{Vafa:1994rv}
C.~Vafa and E.~Witten, \emph{{On Orbifolds with Discrete Torsion}},
  \href{https://doi.org/10.1016/0393-0440(94)00048-9}{\emph{J. Geom. Phys.}
  {\bfseries 15} (1995) 189}
  [\href{https://arxiv.org/abs/hep-th/9409188}{{\ttfamily hep-th/9409188}}].

\bibitem{Lunin:2002bj}
O.~Lunin, S.~D. Mathur and A.~Saxena, \emph{{What is the Gravity Dual of a
  Chiral Primary?}},
  \href{https://doi.org/10.1016/S0550-3213(03)00081-6}{\emph{Nucl. Phys. B}
  {\bfseries 655} (2003) 185}
  [\href{https://arxiv.org/abs/hep-th/0211292}{{\ttfamily hep-th/0211292}}].

\bibitem{Kirby}
R.~C. Kirby, \emph{The topology of 4-manifolds}, vol.~1374. Springer, 2006.

\bibitem{Dijkgraaf:2000fq}
R.~Dijkgraaf, J.~M. Maldacena, G.~W. Moore and E.~P. Verlinde, \emph{{A Black
  Hole Farey Tail}},  \href{https://arxiv.org/abs/hep-th/0005003}{{\ttfamily
  hep-th/0005003}}.

\bibitem{Witten:2007kt}
E.~Witten, \emph{{Three-Dimensional Gravity Revisited}},
  \href{https://arxiv.org/abs/0706.3359}{{\ttfamily 0706.3359}}.

\bibitem{Manschot:2007zb}
J.~Manschot, \emph{{Ad$S^3$ Partition Functions Reconstructed}},
  \href{https://doi.org/10.1088/1126-6708/2007/10/103}{\emph{JHEP} {\bfseries
  10} (2007) 103} [\href{https://arxiv.org/abs/0707.1159}{{\ttfamily
  0707.1159}}].

\bibitem{Manschot:2007ha}
J.~Manschot and G.~W. Moore, \emph{{A Modern Farey Tail}},
  \href{https://doi.org/10.4310/CNTP.2010.v4.n1.a3}{\emph{Commun. Num. Theor.
  Phys.} {\bfseries 4} (2010) 103}
  [\href{https://arxiv.org/abs/0712.0573}{{\ttfamily 0712.0573}}].

\bibitem{Bantay:1998fy}
P.~Bantay, \emph{{Orbifoldization, Covering Surfaces and Uniformization
  Theory}}, \href{https://doi.org/10.1023/A:1017934313775}{\emph{Lett. Math.
  Phys.} {\bfseries 57} (2001) 1}
  [\href{https://arxiv.org/abs/hep-th/9808023}{{\ttfamily hep-th/9808023}}].

\bibitem{Verlinde:2002ig}
H.~L. Verlinde, \emph{{Bits, Matrices and 1/N}},
  \href{https://doi.org/10.1088/1126-6708/2003/12/052}{\emph{JHEP} {\bfseries
  12} (2003) 052} [\href{https://arxiv.org/abs/hep-th/0206059}{{\ttfamily
  hep-th/0206059}}].

\bibitem{Teitelboim:1983ux}
C.~Teitelboim, \emph{{Gravitation and Hamiltonian Structure in Two Space-Time
  Dimensions}}, \href{https://doi.org/10.1016/0370-2693(83)90012-6}{\emph{Phys.
  Lett. B} {\bfseries 126} (1983) 41}.

\bibitem{Jackiw:1984je}
R.~Jackiw, \emph{{Lower Dimensional Gravity}}, .

\bibitem{Liu:1987nz}
J.~Liu and J.~Polchinski, \emph{{Renormalization of the Mobius Volume}},
  \href{https://doi.org/10.1016/0370-2693(88)91566-3}{\emph{Phys. Lett. B}
  {\bfseries 203} (1988) 39}.

\bibitem{Tseytlin:1988tv}
A.~A. Tseytlin, \emph{{Mobius Infinity Subtraction and Effective Action in
  $\sigma$ Model Approach to Closed String Theory}},
  \href{https://doi.org/10.1016/0370-2693(88)90421-2}{\emph{Phys. Lett. B}
  {\bfseries 208} (1988) 221}.

\bibitem{Balthazar:2019rnh}
B.~Balthazar, V.~A. Rodriguez and X.~Yin, \emph{{Zz Instantons and the
  Non-Perturbative Dual of $c = 1$ String Theory}},
  \href{https://arxiv.org/abs/1907.07688}{{\ttfamily 1907.07688}}.

\bibitem{Balthazar:2019ypi}
B.~Balthazar, V.~A. Rodriguez and X.~Yin, \emph{{Multi-Instanton Calculus in $c
  = 1$ String Theory}},  \href{https://arxiv.org/abs/1912.07170}{{\ttfamily
  1912.07170}}.

\bibitem{Sen:2019qqg}
A.~Sen, \emph{{Fixing an Ambiguity in Two Dimensional String Theory Using
  String Field Theory}},
  \href{https://doi.org/10.1007/JHEP03(2020)005}{\emph{JHEP} {\bfseries 03}
  (2020) 005} [\href{https://arxiv.org/abs/1908.02782}{{\ttfamily
  1908.02782}}].

\bibitem{Sen:2020cef}
A.~Sen, \emph{{D-Instanton Perturbation Theory}},
  \href{https://arxiv.org/abs/2002.04043}{{\ttfamily 2002.04043}}.

\bibitem{Li:2020nei}
S.~Li and J.~Troost, \emph{{Twisted String Theory in Anti-de~Sitter Space}},
  \href{https://arxiv.org/abs/2005.13817}{{\ttfamily 2005.13817}}.

\bibitem{Li:2020zwo}
S.~Li and J.~Troost, \emph{{The Topological Symmetric Orbifold}},
  \href{https://arxiv.org/abs/2006.09346}{{\ttfamily 2006.09346}}.

\bibitem{Gukov:2004ym}
S.~Gukov, E.~Martinec, G.~W. Moore and A.~Strominger, \emph{{The Search for a
  holographic dual to ${\rm AdS}_3 \times {\rm S}^3 \times {\rm S}^{3}\times
  {\rm S}^{1}$}}, \href{https://doi.org/10.4310/ATMP.2005.v9.n3.a3,
  10.1142/9789812775344_0035}{\emph{Adv. Theor. Math. Phys.} {\bfseries 9}
  (2005) 435} [\href{https://arxiv.org/abs/hep-th/0403090}{{\ttfamily
  hep-th/0403090}}].

\bibitem{Tong:2014yna}
D.~Tong, \emph{{The holographic dual of ${\rm AdS}_{3} \times {\rm S}^{3}
  \times {\rm S}^{3} \times {\rm S}^{1}$}},
  \href{https://doi.org/10.1007/JHEP04(2014)193}{\emph{JHEP} {\bfseries 04}
  (2014) 193} [\href{https://arxiv.org/abs/1402.5135}{{\ttfamily 1402.5135}}].

\bibitem{Eberhardt:2017pty}
L.~Eberhardt, M.~R. Gaberdiel and W.~Li, \emph{{A holographic dual for string
  theory on ${\rm AdS}_3 \times {\rm S}^{3} \times {\rm S}^{3}\times {\rm
  S}^{1}$}}, \href{https://doi.org/10.1007/JHEP08(2017)111}{\emph{JHEP}
  {\bfseries 08} (2017) 111}
  [\href{https://arxiv.org/abs/1707.02705}{{\ttfamily 1707.02705}}].

\bibitem{Eberhardt:2019niq}
L.~Eberhardt and M.~R. Gaberdiel, \emph{{Strings on $\mathrm{AdS}_3 \times
  \mathrm{S}^3 \times \mathrm{S}^3 \times \mathrm{S}^1$}},
  \href{https://doi.org/10.1007/JHEP06(2019)035}{\emph{JHEP} {\bfseries 06}
  (2019) 035} [\href{https://arxiv.org/abs/1904.01585}{{\ttfamily
  1904.01585}}].

\bibitem{Witten:1998qj}
E.~Witten, \emph{{Anti-de~Sitter Space and Holography}},
  \href{https://doi.org/10.4310/ATMP.1998.v2.n2.a2}{\emph{Adv. Theor. Math.
  Phys.} {\bfseries 2} (1998) 253}
  [\href{https://arxiv.org/abs/hep-th/9802150}{{\ttfamily hep-th/9802150}}].

\bibitem{Maldacena:2004rf}
J.~M. Maldacena and L.~Maoz, \emph{{Wormholes in AdS}},
  \href{https://doi.org/10.1088/1126-6708/2004/02/053}{\emph{JHEP} {\bfseries
  02} (2004) 053} [\href{https://arxiv.org/abs/hep-th/0401024}{{\ttfamily
  hep-th/0401024}}].

\bibitem{Sundborg:1999ue}
B.~Sundborg, \emph{{The Hagedorn transition, deconfinement and N=4 SYM
  theory}}, \href{https://doi.org/10.1016/S0550-3213(00)00044-4}{\emph{Nucl.
  Phys. B} {\bfseries 573} (2000) 349}
  [\href{https://arxiv.org/abs/hep-th/9908001}{{\ttfamily hep-th/9908001}}].

\bibitem{Sundborg:2000wp}
B.~Sundborg, \emph{{Stringy gravity, interacting tensionless strings and
  massless higher spins}},
  \href{https://doi.org/10.1016/S0920-5632(01)01545-6}{\emph{Nucl. Phys. Proc.
  Suppl.} {\bfseries 102} (2001) 113}
  [\href{https://arxiv.org/abs/hep-th/0103247}{{\ttfamily hep-th/0103247}}].

\bibitem{Seiberg:1999xz}
N.~Seiberg and E.~Witten, \emph{{The ${\rm D}1$ / ${\rm D}5$ system and
  singular CFT}},
  \href{https://doi.org/10.1088/1126-6708/1999/04/017}{\emph{JHEP} {\bfseries
  04} (1999) 017} [\href{https://arxiv.org/abs/hep-th/9903224}{{\ttfamily
  hep-th/9903224}}].

\bibitem{Dei:2019osr}
A.~Dei, L.~Eberhardt and M.~R. Gaberdiel, \emph{{Three-point functions in
  $\mathrm{AdS}_{3}/\mathrm{CFT}_{2}$ holography}},
  \href{https://doi.org/10.1007/JHEP12(2019)012}{\emph{JHEP} {\bfseries 12}
  (2019) 012} [\href{https://arxiv.org/abs/1907.13144}{{\ttfamily
  1907.13144}}].

\bibitem{Carson:2014ena}
Z.~Carson, S.~Hampton, S.~D. Mathur and D.~Turton, \emph{{Effect of the
  Deformation Operator in the $\mathrm{D1}-\mathrm{D5}$ CFT}},
  \href{https://doi.org/10.1007/JHEP01(2015)071}{\emph{JHEP} {\bfseries 01}
  (2015) 071} [\href{https://arxiv.org/abs/1410.4543}{{\ttfamily 1410.4543}}].

\bibitem{Larsen:1999uk}
F.~Larsen and E.~J. Martinec, \emph{{$\mathrm{U}(1)$ Charges and Moduli in the
  $\mathrm{D1} - \mathrm{D5}$ System}},
  \href{https://doi.org/10.1088/1126-6708/1999/06/019}{\emph{JHEP} {\bfseries
  06} (1999) 019} [\href{https://arxiv.org/abs/hep-th/9905064}{{\ttfamily
  hep-th/9905064}}].

\bibitem{David:2002wn}
J.~R. David, G.~Mandal and S.~R. Wadia, \emph{{Microscopic formulation of black
  holes in string theory}},
  \href{https://doi.org/10.1016/S0370-1573(02)00271-5}{\emph{Phys. Rept.}
  {\bfseries 369} (2002) 549}
  [\href{https://arxiv.org/abs/hep-th/0203048}{{\ttfamily hep-th/0203048}}].

\bibitem{Skenderis:2009ju}
K.~Skenderis and B.~C. van Rees, \emph{{Holography and Wormholes in 2+1
  Dimensions}}, \href{https://doi.org/10.1007/s00220-010-1163-z}{\emph{Commun.
  Math. Phys.} {\bfseries 301} (2011) 583}
  [\href{https://arxiv.org/abs/0912.2090}{{\ttfamily 0912.2090}}].

\bibitem{Jafferis_talk}
D.~Jafferis, \emph{{Stringy ER=EPR (Talk given KITP conference ``Order from
  Chaos'')}},  {December}, {2018}.

\bibitem{Gaberdiel:2015mra}
M.~R. Gaberdiel and R.~Gopakumar, \emph{{Stringy Symmetries and the Higher Spin
  Square}}, \href{https://doi.org/10.1088/1751-8113/48/18/185402}{\emph{J.
  Phys. A} {\bfseries 48} (2015) 185402}
  [\href{https://arxiv.org/abs/1501.07236}{{\ttfamily 1501.07236}}].

\bibitem{Gaberdiel:2015wpo}
M.~R. Gaberdiel and R.~Gopakumar, \emph{{String Theory as a Higher Spin
  Theory}}, \href{https://doi.org/10.1007/JHEP09(2016)085}{\emph{JHEP}
  {\bfseries 09} (2016) 085}
  [\href{https://arxiv.org/abs/1512.07237}{{\ttfamily 1512.07237}}].

\bibitem{Maldacena:2000hw}
J.~M. Maldacena and H.~Ooguri, \emph{{Strings in ${\rm AdS}_3$ and ${\rm
  SL}(2,\mathds{R})$ WZW model 1.: The Spectrum}},
  \href{https://doi.org/10.1063/1.1377273}{\emph{J. Math. Phys.} {\bfseries 42}
  (2001) 2929} [\href{https://arxiv.org/abs/hep-th/0001053}{{\ttfamily
  hep-th/0001053}}].

\bibitem{Gawedzki:1991yu}
K.~Gawedzki, \emph{{Noncompact WZW Conformal Field Theories}},  in \emph{{Nato
  Advanced Study Institute: New Symmetry Principles in Quantum Field Theory}},
  pp.~0247--274, 10, 1991,
  \href{https://arxiv.org/abs/hep-th/9110076}{{\ttfamily hep-th/9110076}}.

\end{thebibliography}\endgroup
\end{document}